\documentclass[a4paper,10pt]{article}
\usepackage[utf8]{inputenc}

\usepackage{amsmath,amsthm,amssymb,amsfonts,stmaryrd,tensor,scalerel,hyperref}
\usepackage{graphicx}
\usepackage{xcolor}
\usepackage[all,cmtip]{xy}
\usepackage{placeins}
\usepackage{array}
\usepackage{multirow}
\usepackage[affil-it]{authblk}

\usepackage{pgf}
\usepackage[absolute, overlay]{textpos}

\textblockorigin{\paperwidth}{0.0 pt}

\theoremstyle{plain}
\newtheorem{remark}{Remark}[section]
\theoremstyle{definition}
\newtheorem{definition}{Definition}[section]

\newcommand{\ro}{\epsilon}

\newcommand{\fr}{F}

\newcommand{\hess}{\det\textnormal{Hess}}

\newcommand{\rn}[1]{{\mathbb{R}^#1}}

\newcommand{\punto}{\,\cdot\,}

\newcommand{\cotb}[1]{T^*\mathbb{R}^#1}

\newcommand{\MA}{Monge--Ampère }
\newcommand{\LR}{Lychagin--Roubtsov }
\newcommand{\CS}{Chynoweth--Sewell }

\newcommand{\PV}{q_g}

\newcommand{\piL}{\pi_{{}_L}}

\newcommand{\cLt}{{\Check{L}_t}}

\newcommand{\symplectic}{\omega}
\newcommand{\effective}{\alpha}

\newcommand{\scalar}{S\hspace{-1.3pt}c}

\title{\MA geometry and the Eady problem}

\author{R. D'Onofrio${}^{1,3,4}$, G. Ortenzi${}^{2,5}$ and I. Roulstone${}^3$.  }
\date{\today}

\affil{
{\small  $^1$Dipartimento di Matematica e Applicazioni, Universit\`a di  Milano-Bicocca, \\ Via Roberto Cozzi 55, I-20125 Milano, 
Italy}\\
{\small  (r.donofrio1@campus.unimib.it)}
\medskip\\

{\small ${}^2$Dipartimento di Matematica ‘‘Giuseppe Peano'',\\  Via Carlo Alberto 10, 10123 Torino, Italy}\\
{\small (giovanni.ortenzi@unito.it)}
\medskip\\

{\small  $^3$School of Mathematics and Physics, University of Surrey, Guildford GU2 7XH, UK
}\\
{\small  (r.d'onofrio@surrey.ac.uk, i.roulstone@surrey.ac.uk)}
\medskip

{\small ${}^4$INFN, Sezione di Milano-Bicocca, Piazza della Scienza 3, 20126 Milano, Italy}
\medskip\\

{\small ${}^5$INFN, Sezione di Torino, Via Pietro Giuria 1, 10125 Torino, Italy}

}

\date{\today}

\begin{document}

\maketitle

\pgfmathwidth{"top right corner at page \thepage"}
\begin{textblock}{\pgfmathresult}[1, 0](0, 1)
\noindent
{\fontfamily{cmr}\selectfont
DMUS-MP-24/02}
\end{textblock}

\begin{abstract}
    Chynoweth and Sewell proposed a mathematical model for an atmospheric front based on the singularities of the Legendre transformation between different pairs of dual variables.
    Drawing inspiration from their work, we formalize the idea of a \CS front and illuminate its geometrical meaning from the viewpoint of \MA geometry. This extends our previous work on the semigeostrophic model to dynamically-forming singularities.
    We use the notion of a \CS front to characterize the classical Eady problem and its known solutions in geometrical terms.
\end{abstract}

\section{Introduction}
In certain conditions, atmospheric flows can become unstable. One of the most important examples is baroclinic instability, considered the main mechanism leading to the formation of mid-latitude cyclones and weather fronts. 
Typically, baroclinic instability manifests itself when a shear flow in the zonal direction is combined with a meridional temperature gradient.
In such conditions, a small-amplitude perturbation of sufficiently large wavelength can grow exponentially in time. The resulting wave pattern is known as a baroclinic wave and represents the early stages of cyclone formation.

A mathematical model capable of representing baroclinic waves (including frontogenesis) is provided by the semigeostrophic equations.
The first systematic study of the semigeostrophic equations dates back to the seminal work of Hoskins \cite{Hos75}, where the so called ‘‘geostrophic momentum transformation’’ was employed to build a solution representing an unstable baroclinic wave.
The geostrophic momentum transformation was later recognised as a form of Legendre transformation by Chynoweth and Sewell \cite{CS89}. Using this point of view, they were able to model fronts by the swallowtail singularities found in the graph of a multivalued geopotential function obtained through a singular inverse Legendre map.


In this paper, we study the Eady problem from the point of view of \MA geometry, whose implications in the context of semigeostrophic equations has been explored in \cite{DOR2023}.
We show that the Hoskins solution of the Eady problem, for sufficiently large times, produces weather fronts in the Chynoweth and Sewell sense.
The geometric perspective allows us to highlight the aspects of the Eady problem connected with catastrophe theory, and show the relationship between \CS fronts and shocks in 1D gas dynamics.
Furthermore, building on our results in \cite{DOR2023}, we explore the pseudo-Riemannian geometry of Eady waves seen as Lagrangian submanifolds of the phase space.
We show that their curvature depends exclusively on the signature of the pull-back \LR metric on them, a fact that reveals the essentially two-dimensional nature of such solutions.

This article is structured as follows. We review the semigeostrophic equations in Section \ref{sec:SG eq}, and the Eady problem with the Hoskins solution in Section \ref{sec:Eady wave}.
Next, we adopt the geometric perspective in Section \ref{sec:MA geom}. Here, we formalize the notion of a \CS front, which we characterize from the catastrophe theory perspective, and show its similarity to 1D gas dynamics shocks.
The section is closed with a discussion on the physical picture associated with an unstable Eady wave, and the velocity field it induces in the atmosphere.
Section \ref{sec:LR metric} collects the results obtained so far about the curvature of Eady waves in the phase space and its physical meaning.

\section{Background -- semigeostrophic equations}
\label{sec:SG eq}
We consider the incompressible three-dimensional semigeostrophic equations in the form given in \cite{Hos75},
\begin{equation}
\label{SG system}
    \begin{cases}
    \vspace{1mm}
    \displaystyle{\frac{D u_g}{D t}-fv+\frac{\partial \phi}{\partial x}=0,\quad \frac{D v_g}{D t}+fu+\frac{\partial \phi}{\partial y}=0,\quad \frac{g\theta}{\theta_0}=\frac{\partial \phi}{\partial z},} & (momentum)\\
    \vspace{1mm}
    \displaystyle{ \frac{\partial u}{\partial x}+\frac{\partial v}{\partial y}+\frac{\partial w}{\partial z}=0,} & (mass)\\
    \displaystyle{\frac{D\theta}{D t}=0,} & (energy)
    \end{cases}
\end{equation}
where
\begin{equation}
\label{material derivative}
    \frac{D}{Dt}=\frac{\partial}{\partial t}+u\frac{\partial}{\partial x}+v\frac{\partial}{\partial y}+w\frac{\partial}{\partial z},
\end{equation}
is the material time derivative, and $\{x,y,z\}$ is a Cartesian coordinate system in a portion of the atmosphere with $y$ directed poleward and $z$ directed vertically. The unknowns are the fluid velocity $(u,v,w)$, the geopotential $\phi$, and the potential temperature $\theta$. The geostrophic component of the velocity field is related to the geopotential by
\begin{equation}
    \label{geostrophic wind}
    u_g:=-\frac{1}{f}\frac{\partial\phi}{\partial y},\qquad v_g:=\frac{1}{f}\frac{\partial\phi}{\partial x}.
\end{equation}
The physical constants appearing in (\ref{SG system}) and (\ref{geostrophic wind}) are the Coriolis parameter $f$, the gravitational acceleration $g$, and a reference value $\theta_0$ for the potential temperature.

In order to streamline the calculations below, we adopt a dimensionless set of variables, first introduced in \cite{SC87}, which clears the system from all physical constants. Namely, we take $f^{-1},L,f^2L^2/g,f^2L^2,\theta_0$ as scaling factors for time, horizontal length, vertical length, geopotential, and potential temperature respectively, with $L$ being some unspecified length.

The evolutionary system of equations (\ref{SG system}) and (\ref{geostrophic wind}) is classically\footnote{See for example \cite{RN94}.} written (in dimensionless variables) as
\begin{gather}
\label{evol eq}
    \frac{D X}{D t}=u_g,\qquad \frac{D Y}{D t}=v_g,\qquad \frac{D Z}{D t}=0,
\end{gather}
with
\begin{equation}
\label{grad P}
    X:=\frac{\partial P}{\partial x}=x+v_g,\qquad Y:=\frac{\partial P}{\partial y}=y-u_g,\qquad Z:=\frac{\partial P}{\partial z}=\theta,
\end{equation}
and
\begin{equation}
\label{mod geopot}
    P:=\frac{x^2}{2}+\frac{y^2}{2}+\phi.
\end{equation}

The semigeostrophic equations admit a Lagrangian invariant $\PV$ known as the \textit{potential vorticity}, a combination of the gradients of $u_g,v_g$ and $\theta$.
The potential vorticity is conveniently expressed in terms of the modified geopotential (\ref{mod geopot}) by
\begin{equation}
\label{CS P}
    \hess(P)=\PV.
\end{equation}
In some physically relevant cases, $\PV$ is approximately constant across the fluid domain, and can be assumed so in (\ref{CS P}).
Under this hypothesis, (\ref{CS P}) is interpreted as a \MA equation for the unknown geopotential and can in principle be solved, subject to appropriate boundary conditions, to obtain solutions to the full system (\ref{SG system}).

\section{Background -- the Eady model}
\label{sec:Eady wave}
The simplest setting in which baroclinic instability can manifest is the Eady model \cite{Ead47}. The troposphere is modeled as a strip,
\begin{equation}
\label{troposphere}
    \Omega=\{(x,y,z)\in\rn{3}:0<z<H\},
\end{equation}
with rigid and impermeable lids representing the Earth's surface ($z=0$) and the tropopause ($z=H$). Within this domain, the basic (stationary) flow consists of a linear velocity profile in the zonal direction and a linear temperature gradient in the meridional direction. In formulas,
\begin{equation}
\label{basic state}
    u_g=\frac{U}{H}z,\qquad v_g=0,\qquad \theta=\theta_0+\frac{N^2\theta_0}{g}z-\frac{f\theta_0}{g}\frac{U}{H}y,
\end{equation}
where $U$ represents the top-lid value for the zonal wind and $N$ is the constant buoyancy frequency.

\subsection{Dimensionless variables}
Using the dimensionless variables introduced in Section (\ref{sec:SG eq}), we can write Eady's basic state in the form
\begin{equation}
\label{n-d basic st}
    u_g=Fz,\qquad v_g=0,\qquad \theta=1+z-Fy,
\end{equation}
where
\begin{equation}
    F:=\frac{U}{NH}
\end{equation}
is the Froude number \cite{HH2013}, and we have set
\begin{equation}
    N=\frac{g}{fL},
\end{equation}
consistently with the usual definition of the buoyancy frequency (see for example \cite{Hos75}).
Note that, in dimensionless variables, the atmosphere domain becomes
\begin{equation}
    0<z<\frac{gH}{f^2L^2}=B,
\end{equation}
where
\begin{equation}
    B:=\frac{NH}{fL}=\frac{gH}{f^2L^2}
\end{equation}
represents the Burger number \cite{HH2013}. Thus, the top-lid value of the zonal velocity equals
\begin{equation}
    u_g\big|_{z=B}=FB\equiv\epsilon.
\end{equation}
The parameter
\begin{equation}
    \epsilon:=\frac{U}{fL}
\end{equation}
represents the Rossby number \cite{HH2013}. The geopotential function corresponding to the unperturbed state is
\begin{equation}
\label{P0}
    P_0=\frac{x^2}{2}+\frac{y^2}{2}+\frac{z^2}{2}-F y z+z.
\end{equation}
The basic state potential vorticity is found from (\ref{CS P}), and turns out to be constant,
\begin{equation}
\label{PV Eady}
    \PV=1-F^2.
\end{equation}
We remark that $q_g$ is positive for sufficiently low wind speeds.

\subsection{The Eady problem}
The basic state specified by (\ref{P0}) solves a constant coefficients \MA equation (\ref{CS P}) with $\PV$ given by (\ref{PV Eady}). The linear stability analysis of this basic state can be tackled by introducing perturbations of the geopotential field 
\begin{equation}
    P=P_0+\eta P_1,\qquad \eta\ll 1,
\end{equation}
that leave $\PV$ constant (and equal to (\ref{PV Eady})). These perturbations are subject to time-dependent boundary conditions (see \cite{Hos75}),
\begin{equation}
\label{BCs}
    w=0,\quad \frac{D\theta}{Dt}=0 \qquad \textnormal{on}\qquad z=0,B.
\end{equation}
Looking for a function $P_1$ which satisfies the \MA equation (\ref{CS P}) and the boundary conditions (\ref{BCs}) to the first order is called the Eady problem.
As it stands, the Eady problem is fraught with technical difficulties due to the implicit dependence of the boundary conditions on $P$. The issue resides with the fact that the time derivative operator depends on $P$ though the velocity field. 
As we show next, this is solved by Hoskins' ‘geostrophic momentum transformation', which also achieves a simplification of the \MA equation itself.

\subsection{Geostrophic momentum transformation}

As clarified in \cite{CS89}, Hoskins' change of variable is in fact a partial Legendre transform of $P$ with respect to the pair of dual variables $x\leftrightarrow X$ and $y\leftrightarrow Y$. The geopotential function $P$ is associated with a dual potential $S(X,Y,z,t)$ by
\begin{equation}
\label{Legendre tr S}
    S=Xx+Yy-P,
\end{equation}
where $x$ and $y$ are expressed in terms of $X$ and $Y$ through equations (\ref{grad P}).
The dual potential satisfies a different form of the \MA equation (\ref{CS P}),
\begin{equation}
\label{CS S}
    \PV\bigg(\frac{\partial^2S}{\partial X^2}\frac{\partial^2S}{\partial Y^2}-\bigg(\frac{\partial^2S}{\partial X\partial Y}\bigg)^2\bigg)+\frac{\partial^2S}{\partial z^2}=0,
\end{equation}
called the \textit{\CS equation}. Because it achieves a separation between derivatives in the horizontal and vertical directions, the \CS equation is easier to solve for particular solutions than the original equation for $P$. However, the major advantage introduced by the Legendre transformation is a simplification on the boundary conditions. Note in fact that the advective derivative operator (\ref{material derivative}) may be written in $(X,Y,z)$ variables as
\begin{equation}
    \frac{D}{Dt}=\frac{\partial}{\partial t}+u_g\frac{\partial}{\partial X}+v_g\frac{\partial}{\partial Y}+w\frac{\partial}{\partial z},
\end{equation}
where the first two equations in (\ref{evol eq}) are in use. On the other hand, the Legendre transform implies
\begin{equation}
    \label{inv Legendre map}
    x=\frac{\partial S}{\partial X},\qquad y=\frac{\partial S}{\partial Y},\qquad \frac{\partial P}{\partial z}=-\frac{\partial S}{\partial z},
\end{equation}
so the geostrophic wind and the potential temperature can be expressed in terms of $S$ by
\begin{equation}
\label{grad S}
    u_g=\frac{\partial S}{\partial Y}-Y,\qquad v_g=X-\frac{\partial S}{\partial X},\qquad \theta=-\frac{\partial S}{\partial z}.
\end{equation}
Therefore, the boundary conditions (\ref{BCs}) read
\begin{equation}
\label{BCs S}
    \frac{\partial^2S}{\partial z\partial t}+\bigg(\frac{\partial S}{\partial Y}-Y\bigg)\frac{\partial^2S}{\partial X\partial z}+\bigg(X-\frac{\partial S}{\partial X}\bigg)\frac{\partial^2S}{\partial Y\partial z}=0.
\end{equation}
The domain boundaries are not affected by the Legendre transform, so the boundary conditions (\ref{BCs S}) are still understood to hold on $z=0,B$.

The dual geopotential specifying the Eady basic state (\ref{P0}) reads
\begin{equation}
\label{S0}
    S_0=\frac{X^2}{2}+\frac{Y^2}{2}-\PV \frac{z^2}{2}+F Y z-z.
\end{equation}
Thus, the Eady problem in the new set of variables amounts to finding a first order solution
\begin{equation}
\label{perturbation S}
    S=S_0+\eta S_1, \qquad \eta\ll 1,
\end{equation}
to the \CS equation (\ref{CS S}) that satisfies the boundary conditions (\ref{BCs S}), with $\PV$ given by (\ref{PV Eady}).

\subsection{Solution of the Eady problem in dual variables}
In this section we review the solution to the Eady problem as worked out by Hoskins.
The solution process is simpler in a shifted set of variables,
\begin{equation}
    \Tilde{X}=X-\frac{\epsilon}{2}t,\qquad \Tilde{z}=z-\frac{B}{2},
\end{equation}
which compensate for the lack of symmetry in the basic flow and the domain. Note that the structure of equation (\ref{CS S}) is unaffected by this change of variables. 
We look for a solution of the form
\begin{equation}
\label{ansatz sol}
    S=S_0+\eta S_1(\Tilde{X},Y,\Tilde{z},t),\qquad \eta\ll 1.
\end{equation}
After introducing the ansatz (\ref{ansatz sol}), we find that the first order terms in $\eta$ vanish if the perturbation field satisfies
\begin{equation}
\label{eq S1}
    \PV\bigg(\frac{\partial^2S_1}{\partial \tilde X^2}+\frac{\partial^2 S_1}{\partial Y^2}\bigg)+\frac{\partial^2S_1}{\partial \tilde z^2}=0.
\end{equation}
On the other hand, the time-dependent boundary conditions (\ref{BCs S}) are satisfied to the first order in $\eta$ if
\begin{equation}
    \frac{\partial^2S_1}{\partial\tilde z\partial t}+F\frac{\partial S_1}{\partial\tilde X}+\frac{\epsilon}{2}\frac{\partial^2 S_1}{\partial\tilde X\partial \tilde z}=0\quad \textnormal{at}\quad \tilde z=\frac{B}{2},
\end{equation}
and
\begin{equation}
    \frac{\partial^2S_1}{\partial\tilde z\partial t}+F\frac{\partial S_1}{\partial\tilde X}-\frac{\epsilon}{2}\frac{\partial^2 S_1}{\partial\tilde X\partial \tilde z}=0\quad \textnormal{at} \quad \tilde z=-\frac{B}{2}.
\end{equation}
Next, we assume that $S_1$ is a monochromatic wave in $\tilde X$ and $Y$ with $\tilde z$-depending amplitude,
\begin{equation}
\label{ansatz S1}
    S_1=\psi(\tilde z)e^{i(k\tilde X+lY-\omega t)}.
\end{equation}
This perturbation field solves (\ref{eq S1}) if
\begin{equation}
    \psi''=m^2 \PV \psi,
\end{equation}
that is,
\begin{equation}
\label{psi}
    \psi(\tilde z)=C_1 e^{m\sqrt{\PV}\tilde z}+C_2 e^{-m\sqrt{\PV}\tilde z}.
\end{equation}
\begin{remark}
Despite being obtained as an approximate solution to the Eady problem, (\ref{ansatz sol}) together with (\ref{ansatz S1}) and (\ref{psi}), is in fact an exact solution. Higher order terms in $\eta$ happen to cancel out in both equation (\ref{CS S}) and the boundary conditions (\ref{BCs S}).
\end{remark}
Next, we fix the constants of integration in (\ref{psi}) by using the boundary conditions.
The boundary conditions at $z=B$ give
\begin{gather*}
    e^{\frac{1}{2}Bm\sqrt{\PV}}(2Fk+m\sqrt{\PV}(-k\epsilon+2\omega))C_1+\\
    e^{-\frac{1}{2}Bm\sqrt{\PV}}(2Fk+m\sqrt{\PV}(k\epsilon-2\omega))C_2=0,
\end{gather*}
and those at $z=0$ give
\begin{gather*}
    e^{-\frac{1}{2}Bm\sqrt{\PV}}(2Fk+m\sqrt{\PV}(k\epsilon+2\omega))C_1+\\
    e^{\frac{1}{2}Bm\sqrt{\PV}}(2Fk-m\sqrt{\PV}(k\epsilon+2\omega))C_2=0.
\end{gather*}
This system admits nontrivial solutions $C_1,C_2$ if the dispersion relation is satisfied,
\begin{equation}
\label{disp relation}
    -\frac{m^2\PV}{F^2k^2}\omega^2=\bigg(1-\frac{Bm\sqrt{\PV}}{2}\coth\frac{Bm\sqrt{\PV}}{2}\bigg)\bigg(\frac{Bm\sqrt{\PV}}{2}\tanh{\frac{Bm\sqrt{\PV}}{2}}-1\bigg).
\end{equation}
In this case, infinite choices for $C_1,C_2$ are allowed.
A particular choice is
\begin{equation}
\label{C1}
    C_1=e^{\frac{1}{2}Bm\sqrt{\PV}}(2Fk-m\sqrt{\PV}(k\epsilon+2\omega)),
\end{equation}
\begin{equation}
\label{C2}
    C_2=-e^{-\frac{1}{2}Bm\sqrt{\PV}}(2Fk+m\sqrt{\PV}(k\epsilon+2\omega)),
\end{equation}
which automatically satisfies the boundary conditions at $z=0$ (the boundary conditions at $z=B$ are satisfied if the dispersion relation holds).
As we show in Section \ref{subsec:x-waves},
the dispersion relation (\ref{disp relation}) produces both stable and unstable modes, with instability occurring for sufficiently long waves.


\section{Geometric picture and singularities}
\label{sec:MA geom}
Working in two spatial dimensions $(x,z)$, Chynoweth and Sewell \cite{CS89} proposed a mathematical model for fronts based on the Legendre transform. 
This builds on the observation that for a given function $S(X,z)$ the corresponding geopotential $P=Xx-S$ may be multivalued. In specific cases (see also \cite{Kos91}), it is possible to excise a single valued convex function from its graph with discontinuous first derivatives across one or more lines in the $(x,z)$-plane. These singular lines are then interpreted as fronts, as the momentum and temperature fields experience a jump discontinuity across them (cf. equations (\ref{grad P})).

Using the same mathematical ideas, we introduce a notion of three-dimensional fronts by applying Chynoweth and Sewell's reasoning in three spatial dimensions. We use the terminology \textit{\CS fronts} to identify both 2D and 3D fronts built this way.

Hoskins' solution to the Eady problem described in the previous section produces \CS fronts in finite time if an unstable mode is selected (see Section \ref{subsec:x-waves}).
In this case, the perturbation can grow up to a point when the Legendre transform becomes singular and swallowtail singularities occur in the graph of $P$.

In this section, we formalize the notion of \CS fronts using the Legendre transformation. Next, we switch point of view, and rephrase this concept in the language of \MA geometry. This allows us to illuminate the inherent similarity existing between \CS fronts and shocks in 1D gas flows.
We conclude this section with a description of the physical picture represented by an unstable Eady wave.

\subsection{\CS fronts}
\label{subsec:CS fronts}
Chynoweth and Sewell's construction \cite{CS89} of 2D fronts is based on the convexity principle introduced in \cite{CP84}, which requires the convexity of $P$ as a necessary condition for the stability of the solution. 
This provides a criterion to select physically meaningful branches of a multivalued $P$. Specifically, Chynoweth and Sewell showed that when $P$ contains a swallowtail singularity, it is possible to excise a globally continuous function $\check P$ from its graph with discontinuous gradient across a line in the $(x,z)$-plane. This singular line is interpreted as a front because absolute momentum and potential temperature suffer from a jump discontinuity across it (cf. equation (\ref{grad P})).

As discussed in \cite{CS89}, convexity of $P$ carries over in different ways to its Legendre dual potentials. Specifically, a convex $P(x,y,z)$ corresponds to a saddle shaped $S(X,Y,z)$ (convex in $(X,Y)$ and concave in $z$).
This observation allows one to switch viewpoint and operate on $S$ to build a dual potential $\check{S}$ that provides, under the inverse Legendre transform, a single-valued geopotential $\check P$ with the required properties. We apply these ideas to build \CS fronts in three dimensions. Our approach reduces to the original set of equations in \cite{CS89} when a particular class of solutions is considered.

Let $S(X,Y,z)$ be a given regular solution to the \CS equation (\ref{CS S}). For every fixed $z=constant$, we build the convex envelope $\Check{S}(\punto,\punto,z)$ of $S(\punto,\punto,z)$. Therefore, $\Check{S}(X,Y,z)$ is convex in $(X,Y)$ and concave in $z$.
Thus, the partial Legendre transform of $\Check{S}$ yields a convex function $\Check{P}(x,y,z)$. Moreover, if $\Check{S}\ne S$, then $\nabla\Check{P}$ is discontinuous across one or more singular surfaces in $\rn{3}$ which are interpreted as fronts. In other words,

\begin{definition}
\CS fronts are the image, under the Legendre map, of the set in $(X,Y,z)$-space where $S(\punto,\punto,z)$ is concave.
\end{definition}

\begin{remark}
    The convex envelope of a given regular function is obtained as follows. Let $S(X,Y)$ be a given regular function which is everywhere convex except in some bounded domain $D\subset \rn{2}$. Finding its convex envelope $\Check{S}(X,Y)$ amounts to solving a free-boundary value problem where the \MA equation
    \begin{equation}
    \label{convex env 1}
        \frac{\partial^2s}{\partial X^2}\frac{\partial^2s}{\partial Y^2}-\bigg(\frac{\partial^2 s}{\partial X\partial Y}\bigg)^2=0
    \end{equation}
    holds in $D$, and the boundary conditions
    \begin{equation}
    \label{convex env BCs}
        s=S,\qquad \nabla s=\nabla S,
    \end{equation}
    hold on the unknown boundary $\partial D$. Once a solution $s$ is found, the convex envelope is obtained piecewise as
    \begin{equation}
        \Check{S}(X,Y)=
        \begin{cases}
            S(X,Y), & \textnormal{if }(X,Y)\ni D,\\
            s(X,Y), & \textnormal{if }(X,Y)\in D.
        \end{cases}
    \end{equation}
    If the function $S(X,Y)$ is concave in several regions $D_1,...,D_n$, then one has to solve $n$ free-boundary value problems similar to the one just described in order to build the convex envelope of $S$.
\end{remark}

Finding the convex envelope of $S(\punto,\punto,z)$ is much simpler if the function $S$ has a trivial dependence on the variable $Y$.
As shown in the next section, this precisely occurs when plane Eady waves directed along the $X$-axis are considered. 
Specifically, we require that $\partial_X\partial_YS=0$ and $\partial_Y^2S=1$. This implies that $S$ satisfies a 2D version of the \CS equation,
\begin{equation}
\label{CS S 2d}
    \PV\frac{\partial^2S}{\partial X^2}+\frac{\partial^2S}{\partial z^2}=0.
\end{equation}
A particular class of solutions (containing the $X$-travelling Eady waves), which we call \textit{cylindrical}, is provided by
\begin{equation}
\label{solution class}
    S=\frac{Y^2}{2}+CYz+S'(X,z),\qquad C\in\mathbb R,
\end{equation}
where $S'$ solves (\ref{CS S 2d}).

For this class of solutions, equation (\ref{convex env 1}) implies that $s$ is affine in $x$, whereas the second of (\ref{convex env BCs}) reduces to $\partial_Xs=\partial_XS$.  
The convex envelope problem thus boils down to the set of equations used in \cite{CS89}.
Namely, suppose that, for a fixed $z$, there exists a region $X_1(z)\le X\le X_2(z)$ where $S'(\punto,z)$ is concave. Then, the boundary points $X_1(z),X_2(z)$ are found from the system
\begin{equation}
\label{X1 X2 system}
    \frac{\partial S'}{\partial X}\bigg|_{X_1}=\frac{S'|_{X_2}-S'|_{X_1}}{X_2-X_1},\qquad \frac{\partial S'}{\partial X}\bigg|_{X_2}=\frac{S'|_{X_2}-S'|_{X_1}}{X_2-X_1},
\end{equation}
and the convex envelope $\Check{S}'(\punto,z)$ is obtained as
\begin{equation}
    \Check{S}'(X,z)=\begin{cases}
        S'(X,z), & \textnormal{if }X<X_1(z)\textnormal{ or }X>X_2(z),\\
        S'(X_1,z)+\frac{\partial S'}{\partial X}(X_1,z)(X-X_1), & \textnormal{if }X_1(z)<X<X_2(z).
    \end{cases}
\end{equation}

For the class of solutions (\ref{solution class}), the domain on $(X,Y,z)$-space where $S$ is concave has always the form
\begin{equation}
    \bigcup_z [X_1(z),X_2(z)]\times\mathbb R,
\end{equation}
($\mathbb R$ stands for the $Y$-axis). The image of this 3D domain under the inverse Legendre transform gives a 2D surface that represent a \CS front in the physical space. A front obtained this way is always a cylindrical surface, meaning that each section $y=constant$ looks the same.

In the following two sections, we recall the basics about \MA geometry, which we use in Section \ref{sec:MA&fronts} to characterize \CS fronts from a novel perspective.

\subsection{Geometry of \MA equations}
A \MA equation is a second order nonlinear PDE where the highest order derivatives appear as combination of the minors of the unknown's Hessian matrix. A symplectic \MA equation is one whose coefficients do not depend on the dependent variable (but can depend on its gradient). Equation (\ref{CS P}) is an example of a symplectic \MA equation provided that $\PV$ is a given constant. The study of symplectic \MA equations reduces to the geometry of a symplectic manifold, the phase space, endowed with a \MA structure. The phase space represents the cotangent bundle to the manifold of independent variables, and, for equation (\ref{CS P}), is $\cotb{3}$. A \MA structure is a pair of differential forms on $\cotb{3}$ which represents the \MA equation. If we endow $\cotb{3}$ with coordinates $(x,y,z,X,Y,Z)$, the \MA structure associated with equation (\ref{CS P}) is
\begin{equation}
\label{MA structure}
\begin{cases}
    \symplectic=d X\wedge d x+d Y\wedge d y+d Z\wedge d z,\\
    \effective=dX\wedge dY\wedge dZ-\PV dx\wedge dy \wedge dz.
\end{cases}
\end{equation}
The \MA structure has the meaning of a geometrical counterpart of the \MA equation. This is clear once we note that (\ref{MA structure}) vanishes on the graph of the gradient $(X,Y,Z)=\nabla P$ of a function $P$ if and only if $P$ solves (\ref{CS P}). 

One of the major advantages brought by the geometrical approach is that it enables a wider definition of a solution to the \MA equation including singular and multivalued ones. We define a generalized solution to (\ref{CS P}) as a Lagrangian submanifold $L\subset\cotb{3}$ such that $\effective|_L=0$. Lagrangian submanifolds can be locally represented in terms of a single function called a generating function. It can be easily seen that each of the partial Legendre transformations of $P$ is in fact a generating function. For example, a function $S(X,Y,z)$ generates a Lagrangian submanifold as the combined zero set
\begin{equation}
\label{Lagr S}
    L=\bigg\{(x,y,z,X,Y,Z):x=\frac{\partial S}{\partial X},y=\frac{\partial S}{\partial Y},Z=-\frac{\partial S}{\partial z}\bigg\}.
\end{equation}
It is straightforward to check that the condition $\alpha|_L=0$ in this case boils down to the \CS equation (\ref{CS S}).

\subsection{Refresh on catastrophe theory}
The geometrical point of view introduced in the previous section proves especially useful for the classification of singularities. In this framework, every loss of regularity in a (weak) solution to (\ref{CS P}) is interpreted as some kind of projection singularity of $L$ onto the base manifold of the cotangent bundle $\cotb{3}$. This approach, which dates back to the work of Vinogradov and Kupershmidt \cite{VK77}, connects the theory differential equations with catastrophe theory \cite{Lyc85} and enables the use of the theory of Lagrangian singularities (see \cite{Arn78,AGV2012}) for the classification of solutions to \MA equations. To fix ideas, we introduce the bundle projection
\begin{equation}
    \pi:\cotb{3}\to \rn{3}
\end{equation}
as
\begin{equation}
    \pi:(x,y,z,X,Y,Z)\to(x,y,z).
\end{equation}
The restriction $\piL$ of this map to a generalized solution $L$ is singular at those points for which is impossible to find a neighbourhood that can be represented as the graph of a gradient $\nabla P$. The set of such points is denoted $\Sigma L$, and is characterized as
\begin{equation}
    \Sigma L=\{a\in L:\det(d\piL)|_a=0\}.
\end{equation}

If a Lagrangian submanifold is described in terms of a generating function $S(X,Y,z)$ as in (\ref{Lagr S}), the restriction of $\pi$ to $L$ takes the form
\begin{equation}
\label{piL}
    \piL(X,Y,z)=\bigg(\frac{\partial S}{\partial X},\frac{\partial S}{\partial Y},z\bigg).
\end{equation}
In this case, the singular locus can be expressed as
\begin{equation}
\label{def f}
    \Sigma L=\bigg\{(X,Y,z)\in L:f:=\frac{\partial^2S}{\partial X^2}\frac{\partial^2S}{\partial Y^2}-\bigg(\frac{\partial^2 S}{\partial X\partial Y}\bigg)^2=0\bigg\}.
\end{equation}
A classification of singular points can be obtained by looking more closely at the properties of $d\piL$ (see Section I.2 of \cite{AGV2012}). Suppose that the rank of the map $\piL$ drops by just 1 on every point of $\Sigma L$. In this situation, cusp points ($A_3$) are distinguished by fold points ($A_2$) by the condition that the kernel of $d\piL$ is tangent to $\Sigma L$ at the first ones. 
In other words, $A_3$ points are characterized as those points on $\Sigma L$ where the intersection
\begin{equation}
\label{swallowtail points}
    \ker(d\piL)\cap\ker(df)
\end{equation}
contains nonzero vectors.
To express this condition more explicitly, note that $\ker(d\piL)$ is always 1-dimensional on $\Sigma L$, and either
\begin{equation}
\label{ker dpi 1}
    \ker(d\piL)=\bigg\langle \frac{\partial^2 S}{\partial X\partial Y}\partial_X-\frac{\partial^2 S}{\partial X^2}\partial_Y\bigg\rangle,
\end{equation}
or
\begin{equation}
\label{ker dpi 2}
    \ker(d\piL)=\bigg\langle \frac{\partial^2S}{\partial Y^2}\partial_X-\frac{\partial^2S}{\partial Y\partial X}\partial_Y\bigg\rangle.
\end{equation}
Thus we can modify (\ref{swallowtail points}) as
\begin{equation}
\label{swallowtail points 2}
    \ker(d\piL)\subset\ker(df),
\end{equation}
and write it explicitly as either
\begin{equation}
\label{SWT points 1}
    df\bigg(\frac{\partial^2 S}{\partial X\partial Y}\partial_X-\frac{\partial^2 S}{\partial X^2}\partial_Y\bigg)=\frac{\partial f}{\partial X}\frac{\partial^2 S}{\partial X\partial Y}-\frac{\partial f}{\partial Y}\frac{\partial^2S}{\partial X^2}=0,
\end{equation}
if (\ref{ker dpi 1}) holds, or,
\begin{equation}
\label{SWT points 2}
    df\bigg(\frac{\partial^2S}{\partial Y^2}\partial_X-\frac{\partial^2S}{\partial Y\partial X}\partial_Y\bigg)=\frac{\partial f}{\partial X}\frac{\partial^2 S}{\partial Y^2}-\frac{\partial f}{\partial Y}\frac{\partial^2S}{\partial Y\partial X}=0.
\end{equation}
if (\ref{ker dpi 2}) does. This condition, complemented with $f=0$, identifies $A_3$ points on $L$.

\subsection{\MA geometry and fronts}
\label{sec:MA&fronts}

\CS fronts (cf. Section \ref{subsec:CS fronts}) have a natural interpretation in terms of \MA geometry. To see this, we have to move our attention from the graph of $P$ to that of $\nabla P$, i.e., the Lagrangian submanifold $L\subset\cotb{3}$ generated by $P$. 
A basic result from catastrophe theory says that if $P$ has a swallowtail singularity, then its gradient has a cusp singularity (see for example \cite{Sew87}). Being related this way, these classes of singularities are both identified in Arnold's classification by the label $A_3$
\footnote{The swallowtail singularity in the graph of $P$ is a \textit{Legendrian} $A_3$ singularity, whereas the cusp in the graph of $\nabla P$ is a \textit{Lagrangian} $A_3$ singularity.}. 
In the geometrical view, Chynoweth and Sewell's construction of fronts boils down to cutting the Lagrangian submanifold $L$ along a suitably chosen surface and discarding the multivalued piece. This yields a modified Lagrangian submanifold $\Check{L}$ whose projection onto the physical space is everywhere single-valued. The precise geometry of the cutting surface is determined by the Chynoweth--Sewell criterion, as show next in connection with the class of 2-dimensional solutions introduced in (\ref{solution class}). Note first that equations (\ref{X1 X2 system}) can be written as
\begin{equation}
    S'|_{X_2}-S'|_{X_1}=x(X_2-X_1),\qquad x=\frac{\partial S'}{\partial X}\bigg|_{X_1}=\frac{\partial S'}{\partial X}\bigg|_{X_2}.
\end{equation}
On the other hand, a straightforward application of the integration by parts rule yields
\begin{equation}
    S'|_{X_2}-S'|_{X_1}-x(X_2-X_1)=-\int_\gamma Xdx,
\end{equation}
where $\gamma$ is the contour represented in Figure \ref{fig:z-slice}. Thus, the cutting surface is $\gamma\times\mathbb R\subset L$, where the contour $\gamma$ is found from
\begin{equation}
    \int_\gamma Xdx=0.
\end{equation}
This condition can be made more explicit by observing that, by Stokes' theorem,
\begin{equation}
\label{equal area}
    0=\int_{\Sigma_+\cup\Sigma_-} dX\wedge dx=\int_{\Sigma_+}dX\wedge dx-\int_{\Sigma_-}dX\wedge dx,
\end{equation}
where $\Sigma_+,\Sigma_-$ are the regions depicted in Figure \ref{fig:z-slice}. Equation (\ref{equal area}) says that, for each $z=const.$ (and $Y=const.$) slice of $L$, the two regions $\Sigma_+,\Sigma_-$ have equal area. Figure \ref{fig:3d view} provides a schematic depiction of $L$ and $\Check{L}$.

\begin{remark}
Equation (\ref{equal area}) is nothing but a statement of the classical Maxwell rule of shock theory in the context of semigeostrophic equations. This observation in turn suggests that \CS fronts, just as shocks, satisfy a conservation principle. 
Indeed, semigeostrophic flows obey the thermal wind balance, which can be expressed in the form of a pair of conservation laws,
\begin{equation}
\label{TWB}
    \frac{\partial X}{\partial z}=\frac{\partial Z}{\partial x},\qquad \frac{\partial Y}{\partial z}=\frac{\partial Z}{\partial y}.
\end{equation}
From the mathematical viewpoint, these equations are just the compatibility condition which ensure that $(X,Y,Z)=\nabla P$ for some $P$. In the language of \MA geometry, a conservation law is a $(n-1)$-form on $\cotb{n}$ whose pull-back to a solution is closed \cite{Lyc85,KLR2006}. The $2$-forms corresponding to (\ref{TWB}) are, respectively,
\begin{equation}
    \beta_1=(Xdx+Zdz)\wedge dy,\qquad \beta_2=dx\wedge(Ydy+Zdz).
\end{equation}
When cylindrical solutions (\ref{solution class}) are considered, the equal area rule (\ref{equal area}) can be used to derive a Rankine-Hugoniot condition corresponding to the conservation law $\beta_1$. First, note that for cylindrical solutions we always have $Y=y-Cz$. As a consequence, $(X,y,z)$ can be used as coordinates on $L$ in place of $(X,Y,z)$. Next, consider the cylindrical surface in $L$ given by
\begin{equation}
    \Gamma\times [0,1],
\end{equation}
where $\Gamma$ is a certain closed curve in the $(X,z)$ plane and $y\in[0,1]$. By Stokes theorem, we have
\begin{equation}
    \int_{\Gamma\times [0,1]}\beta_1=0.
\end{equation}
This implies
\begin{equation}
    \oint_\Gamma Xdx+Zdz=0.
\end{equation}
Next, suppose that $\Gamma=\gamma\cup\Gamma_+\cup\Gamma_-$ as in Figure \ref{fig:3d view}. Then, the above integral splits into three contributions,
\begin{equation}
\label{3 ints}
    \int_\gamma (Xdx+Zdz)+\int_{\Gamma_+} (Xdx+Zdz)+\int_{\Gamma_-} (Xdx+Zdz)=0.
\end{equation}
Now, the first integral vanishes because of the equal area rule and the fact that $dz=0$ on $\gamma$. Accounting that $\Gamma_+$ and $\Gamma_-$ have the same projection onto the $(x,z)$-plane, we can write (\ref{3 ints}) as a single integral using $z$ as parameter,
\begin{equation}
    \int_{z_1}^{z_2}\bigg(\llbracket X\rrbracket \frac{dx}{dz}+\llbracket Z\rrbracket\bigg)dz=0,
\end{equation}
where $z_1,z_2$ represent the values of $z$ at the two ends of $\Gamma_+$ (or $\Gamma_-$) and $\llbracket\cdot\rrbracket$ represents the classical jump operator. Since at least one among $z_1,z_2$ can be chosen arbitrarily, the previous equation implies the Rankine-Hugoniot condition,
\begin{equation}
    \frac{dx}{dz}=-\frac{\llbracket Z\rrbracket}{\llbracket X\rrbracket}.
\end{equation}
Note that this equation occurs in the papers \cite{CP84,CNP87} where its origin is traced back to the work of Margules.
\begin{figure}
    \centering
    \includegraphics[width=0.6\textwidth]{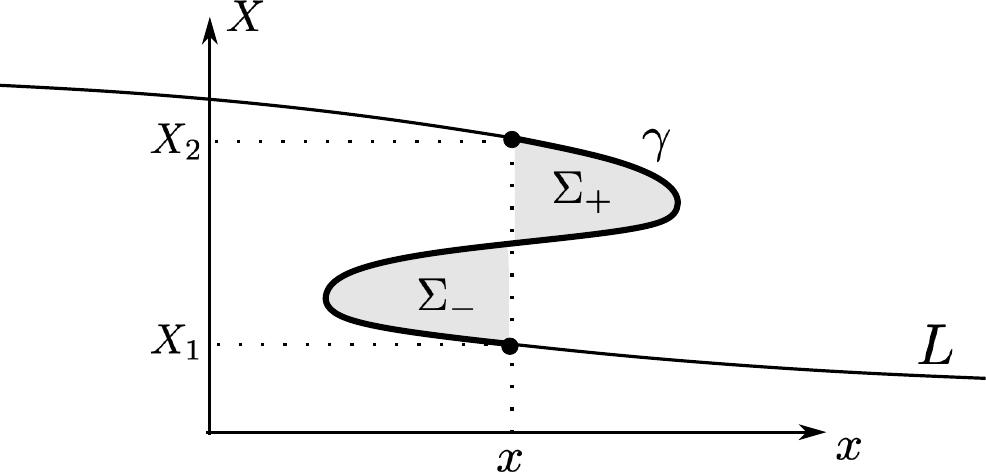}
    \caption{A slice through $L$ for constant $z$ and $y$. The position $x$ at which the front forms is where the areas of $\Sigma_+$ and $\Sigma_-$ coincide. The curve $\gamma$ has end points $(x,X_1)$ and $(x,X_2)$.}
    \label{fig:z-slice}
\end{figure}
\begin{figure}
    \centering
    \includegraphics[width=0.45\textwidth]{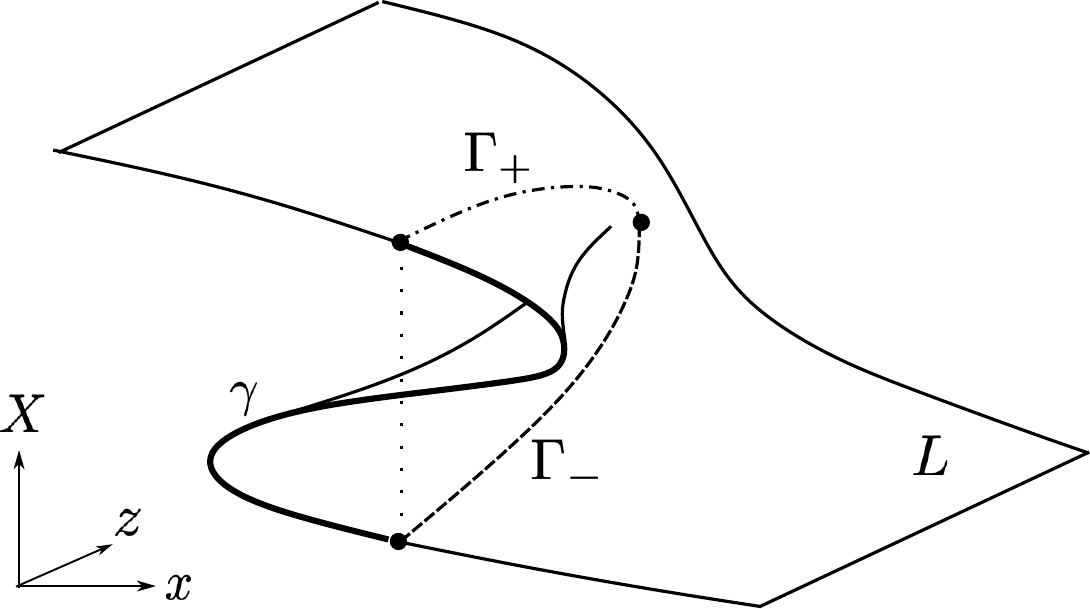}
    \includegraphics[width=0.45\textwidth]{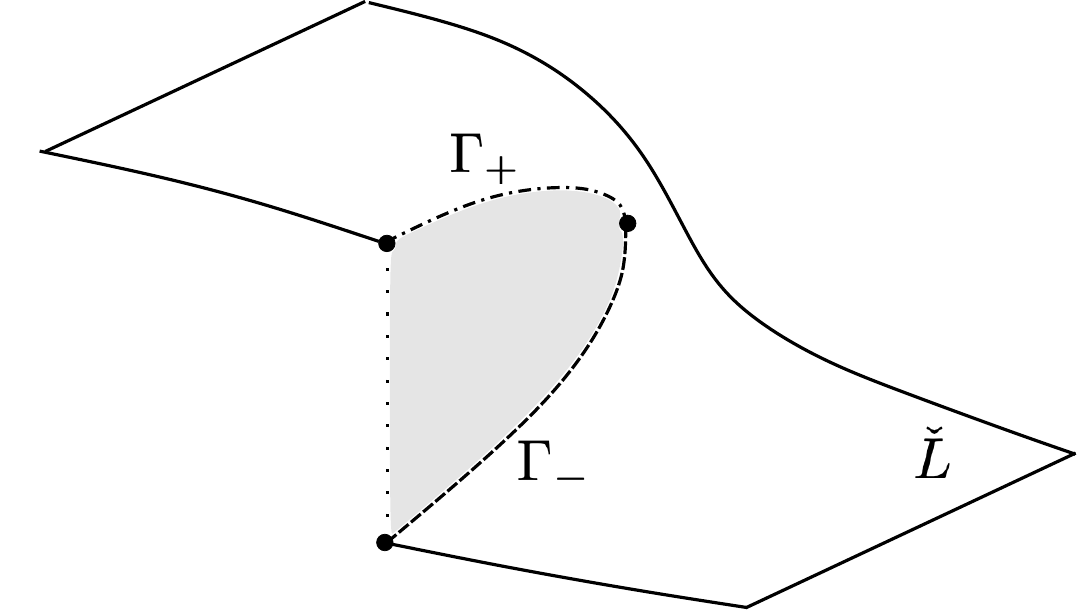}
    \caption{A slice through $L$ for constant $y$. The three curves $\gamma,\Gamma_+,\Gamma_-$ form a closed loop. The curves $\Gamma_+,\Gamma_-$ have the same projection onto the $(x,z)$-plane.}
    \label{fig:3d view}
\end{figure}
\end{remark}

\subsection{Unidirectional waves}
\label{subsec:x-waves}

\begin{figure}
    \centering
    \includegraphics[width=0.5\textwidth]{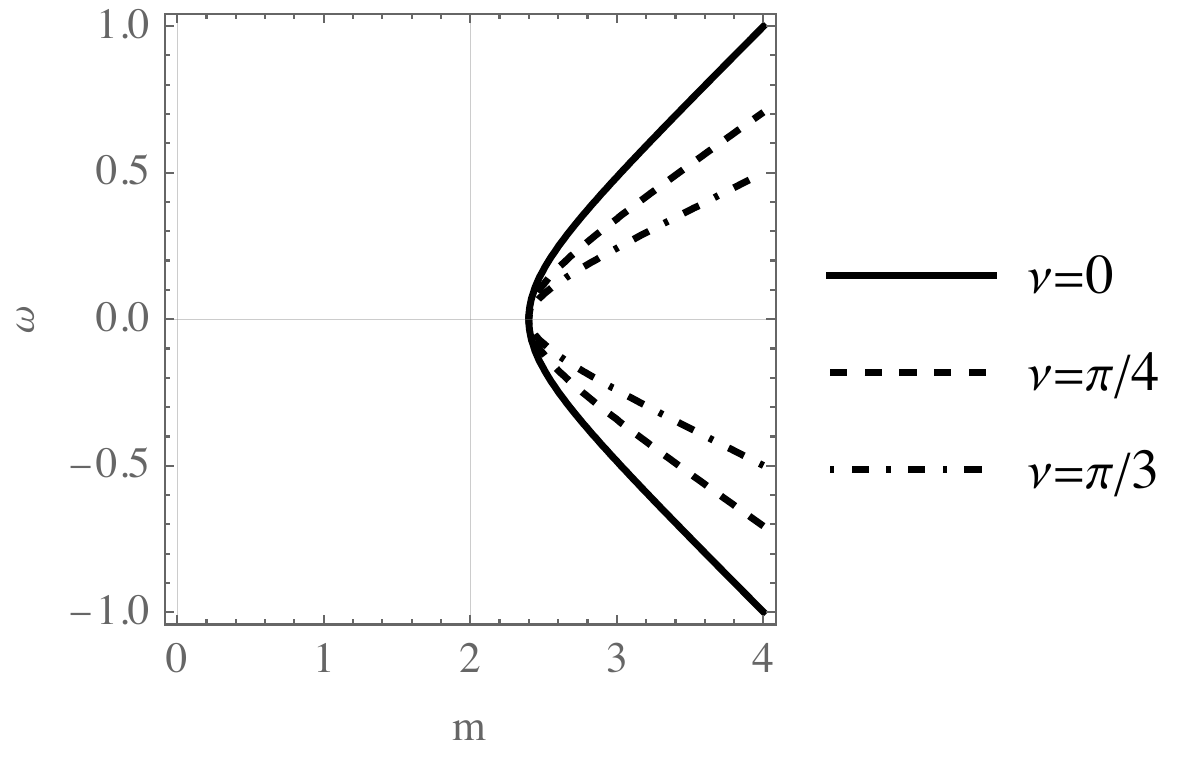}
    \caption{Dispersion relation (\ref{disp relation 2}) with $B=\sqrt{2},F=1/\sqrt{2}$ for different values of the angle $\nu$ that the wave direction forms with the $X$-axis.}
    \label{fig:disp rel}
\end{figure}

The dispersion relation (\ref{disp relation}) can always be written in the form
\begin{equation}
\label{disp relation 2}
    -\omega^2=\cos(\nu)^2\varphi(m;F,B),
\end{equation}
where $\nu=\arctan(l/k)$, and
\begin{equation}
    \varphi(m;F,B)=\frac{F^2}{\PV}\bigg(1-\frac{Bm\sqrt{\PV}}{2}\coth\frac{Bm\sqrt{\PV}}{2}\bigg)\bigg(\frac{Bm\sqrt{\PV}}{2}\tanh{\frac{Bm\sqrt{\PV}}{2}}-1\bigg).
\end{equation}
Equation (\ref{disp relation 2}) says that the qualitative properties of the dispersion relation are unaffected by the direction of propagation of waves (with the only exception being $\nu=\pi/2$). In particular, the magnitude $m$ of the least stable wavenumber is found from
\begin{equation}
    \varphi(m;F,B)=0,
\end{equation}
and only depends on the parameters $F,B$. A physically reasonable choice for these coefficients is
\begin{equation}
\label{coeff choice}
    F=\frac{1}{\sqrt{2}},\quad B=\sqrt{2}\qquad \Rightarrow \qquad \epsilon\equiv FB=1,\quad \PV\equiv 1-F^2=\frac{1}{2},
\end{equation}
for which we obtain,
\begin{equation}
    \varphi(m;1/\sqrt{2},\sqrt{2})=\bigg(1-\frac{m}{2}\coth\frac{m}{2}\bigg)\bigg(\frac{m}{2}\tanh{\frac{m}{2}}-1\bigg).
\end{equation}
The corresponding dispersion relation (\ref{disp relation 2}) is depicted in Figure \ref{fig:disp rel} for different values of $\nu$. In order to discuss the qualitative properties of an unstable Eady wave, we fix $m=2$, which corresponds to the purely imaginary frequency
\begin{equation}
    \omega=\frac{2i\cos(\nu)}{\sqrt{e^4-1}}.
\end{equation}
The simplest solution having $m=2$ is obtained by setting $k=2,l=0$ (which implies $\nu=0$) and represents an $X$-travelling wave,
\begin{gather}
\begin{split}
\label{solution}
    S=\frac{X^2}{2}+\frac{Y^2}{2}-\frac{z^2}{4}+\frac{Yz}{\sqrt{2}}-z
    -2\sqrt{2}\eta e^{\omega_i t}\bigg[2e^{-\sqrt{2}z}\cos(t-2X) + \\ + \omega_i(e^{\sqrt{2}z}+e^{-\sqrt{2}z})\sin(t-2X)\bigg],
\end{split}
\end{gather}
where
\begin{equation}
    \omega_i=\frac{2}{\sqrt{e^4-1}}.
\end{equation}

\begin{figure}
    \centering
    \includegraphics[width=\textwidth]{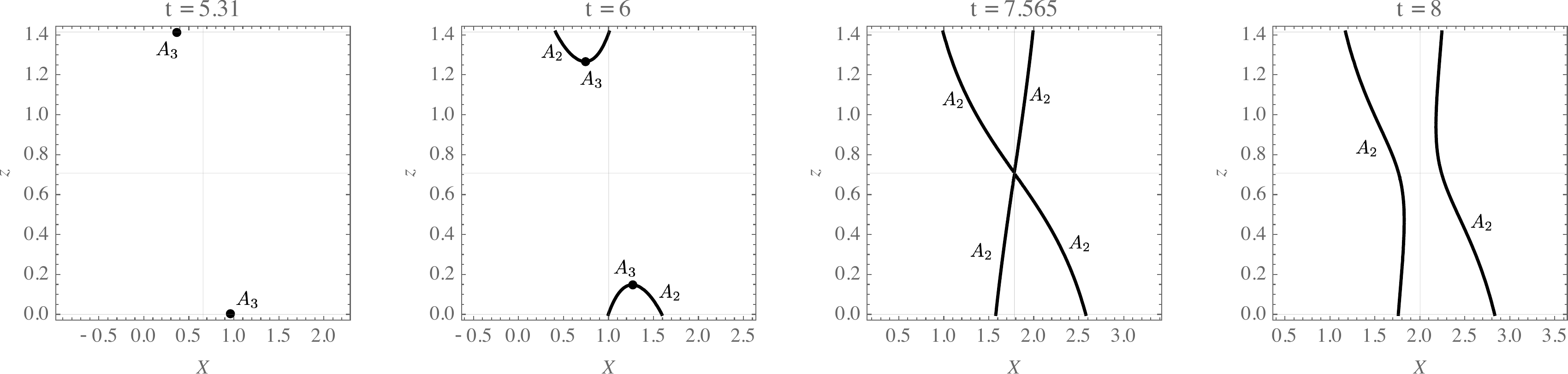}
    \caption{A slice through the singular locus $\Sigma L_t$ for constant $Y$ at different times.}
    \label{fig:sigma L}
\end{figure}

Note that (\ref{solution}) provides an example of a cylindrical solution in the sense of (\ref{solution class}) if the time $t$ is understood as a fixed parameter. Equation (\ref{solution}) may be seen as a $1$-parameter family of generating functions which describes a family of Lagrangian submanifolds $L_t$. Not every $L_t$ has nice projection to the physical space (singularities may appear in $L_t$ as $t$ varies). The singular locus of $L_t$ is
\begin{equation}
    \Sigma L_t=\bigg\{(X,Y,z)\in L_t:\,f(X,z,t)\equiv\frac{\partial^2S}{\partial X^2}=0\,,\,0<z<\sqrt{2}\bigg\},
\end{equation}
and is independent on $Y$. Figure \ref{fig:sigma L} depicts a slice
\begin{equation}
    \Sigma L_t\cap \{Y=const.\},
\end{equation}
through the singular locus, and shows that two topological changes happen over the time evolution (at $t=t'$ and $t=t''$).

The first one occurs at $t=t'$, which is the least time for which $\Sigma L_t\ne \varnothing$. At this time, the singular set (over one period of the solution) is made of two $A_3$ points at the domain boundaries $z=0$ and $z=\sqrt{2}$ respectively.
For $t'<t<t''$, the singular locus is a disconnected set made of two surfaces (curves in the $(X,z)$-plane) which originate and terminate on either $z=0$ or $z=\sqrt{2}$. Each of these curves comprise a cusp point $A_3$ and a continuous set of fold points $A_2$. These two curves meet at $t=t''$, when the two $A_3$ points coalesce, and a higher order singularity is produced. For $t>t''$, the singular set is made of two disconnected components each of which comprises fold points ($A_2$) only.

\begin{figure}
    \centering
    \includegraphics[width=0.24\textwidth]{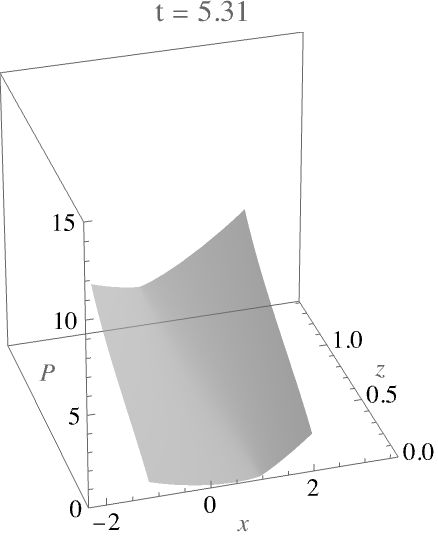}
    \includegraphics[width=0.24\textwidth]{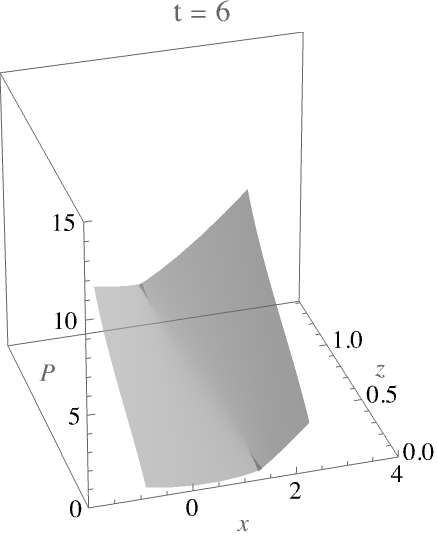}
    \includegraphics[width=0.24\textwidth]{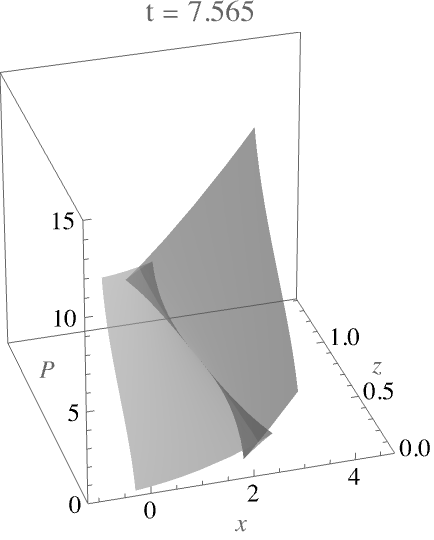}
    \includegraphics[width=0.24\textwidth]{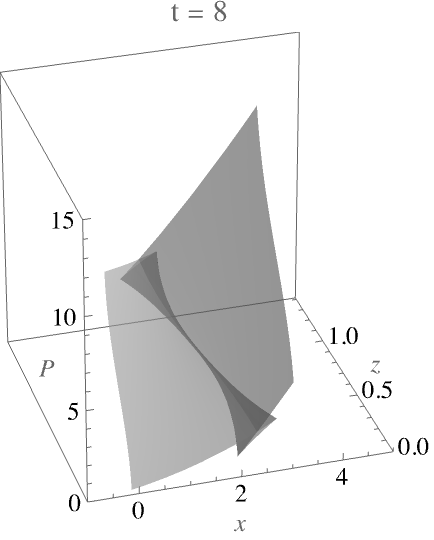}
    \caption{The slice $y=0$ trough the graph of the multivalued $P$ obtained by Legendre transforming (\ref{solution}). The snapshots refer to times $t=t',6.5,t'',8.5$ (cf. (\ref{t'}), (\ref{t''})).}
    \label{fig:P swallowtail}
\end{figure}

Figure \ref{fig:P swallowtail} shows the effects of this time evolution on the graph of $P$. The whole process can be summarized as follows. A pair of swallowtail singularities appear at $t=t'$ in the graph of $P$; they keep growing until they merge at $t=t''$; the two swallowtail points ($A_3$) are annihilated at $t=t''$, and two cusped edges plus a self-intersection line are left for $t>t''$.

The two singular times $t',t''$ are determined as follows. First note that equation (\ref{SWT points 1}) is identically satisfied because the vector field
\begin{equation}
    \frac{\partial^2S}{\partial X\partial Y}\partial_X-\frac{\partial^2S}{\partial X^2}\partial_Y
\end{equation}
vanishes identically on $\Sigma L_t$.
On the other hand, equation (\ref{SWT points 2}) boils down to $\partial_Xf=0$. Therefore, the set of $A_3$ points in $\Sigma L_t$ is implicitly defined by
\begin{equation}
    \begin{cases}
        f(X,z,t)=0,\\
        \frac{\partial f}{\partial X}(X,z,t)=0,
    \end{cases}
\end{equation}
where $f=\partial^2_XS$ and $t$ is fixed. If we set $z=0$ or $z=\sqrt{2}$, we obtain the least catastrophe time as
\begin{equation}
\label{t'}
    t'\approx 5.3,
\end{equation}
which corresponds to the formation of the $A_3$ singularities at the boundary $z=0$ of $L$ at the positions
\begin{equation}
    X\approx 4.1+k\pi\quad (k\in\mathbb{Z}).
\end{equation}
Setting $z=\sqrt{2}/2$ yields the second catastrophe time,
\begin{equation}
\label{t''}
    t''\approx 7.6,
\end{equation}
and
\begin{equation}
    X\approx 4.9+k\pi.
\end{equation}

\subsection{Reconstruction of the velocity field}
In order to understand the physical meaning of an Eady wave it is helpful to consider the velocity field $\mathbf u=(u,v,w)$ it induces in the atmosphere.
By definition, the velocity field on $\rn{3}$ is
\begin{equation}
    u=\frac{Dx}{Dt},\qquad v=\frac{Dy}{Dt},\qquad w=\frac{Dz}{Dt}.
\end{equation}
The position $(x,y,z)$ of the fluid particles is provided by the (time-dependent) projection mapping $\piL:L_t\to\rn{3}$, which, for $L_t$ generated by $S$ reads
\begin{equation}
    \pi(X,Y,z)=(\partial_XS,\partial_YS,z).
\end{equation}
Recalling that, by definition,
\begin{equation}
    u\equiv \frac{Dx}{Dt},\qquad v\equiv \frac{Dy}{Dt}, \qquad \bigg(\textnormal{and }w\equiv \frac{Dz}{Dt}\bigg),
\end{equation}
we may write
\begin{equation}
\label{u(S)}
    u=\frac{D(\partial_XS)}{Dt}=\frac{\partial^2S}{\partial t\partial X}+u_g\frac{\partial^2S}{\partial X^2}+v_g\frac{\partial^2S}{\partial X\partial Y}+w\frac{\partial^2S}{\partial X\partial z},
\end{equation}
\begin{equation}
\label{v(S)}
    v=\frac{D(\partial_YS)}{Dt}
    =\frac{\partial^2S}{\partial t\partial Y}+u_g\frac{\partial^2S}{\partial X\partial Y}+v_g\frac{\partial^2S}{\partial Y^2}+w\frac{\partial^2S}{\partial Y\partial z}.
\end{equation}
On the other hand, $w$ is directly provided by the isentropic flow condition $D\theta/Dt=0$, written as
\begin{equation}
\label{w(S)}
    w=-\frac{\frac{\partial^2S}{\partial z\partial t}+\left(\frac{\partial S}{\partial Y}-Y\right)\frac{\partial^2S}{\partial X\partial z}+\left(X-\frac{\partial S}{\partial X}\right)\frac{\partial^2S}{\partial Y\partial z}}{\frac{\partial^2S}{\partial z^2}}.
\end{equation}
Finally, the geostrophic wind $(u_g,v_g)$ is expressed in terms of $S$ by (\ref{grad S}), i.e.,
\begin{equation}
\label{geos wind S}
    u_g=\frac{\partial S}{\partial Y}-Y,\qquad v_g=X-\frac{\partial S}{\partial X}.
\end{equation}
Equations (\ref{u(S)}), (\ref{v(S)}), and (\ref{w(S)}), together with (\ref{geos wind S}), provide a parametric representation of the velocity field induced on $\rn{3}$ by the generalized solution $L_t$.

We remark that the velocity $\mathbf u$ obtained by this algorithm is generally multivalued. Single valuedness is only ensured when $L_t$ is a classical solution.
For example, if $L_t$ contains a cusp ($A_3$ point), then $\mathbf u$ will be three-valued somewhere in its domain. Single-valuedness may be restored by introducing a \CS front, which formally amounts to restricting the projection mapping $\pi$ to the modified Lagrangian submanifold $\cLt$ when computing (\ref{u(S)}), (\ref{v(S)}), and (\ref{w(S)}).

As the solution (\ref{solution}) is a travelling periodic wave in the $X$-direction, it is convenient, for plotting purposes, to fix the attention on the moving observation window,
\begin{equation}
\label{period}
    X_c(t)-\pi/2<X<X_c(t)+\pi/2,
\end{equation}
which advances linearly in time,
\begin{equation}
    X_c(t):=X_0+\frac{t-t_0}{2},
\end{equation}
and it is centered upon a developing front by suitably choosing the coefficients $(X_0,t_0)$. 

\begin{figure}
    \centering
    \includegraphics[width=0.8\textwidth]{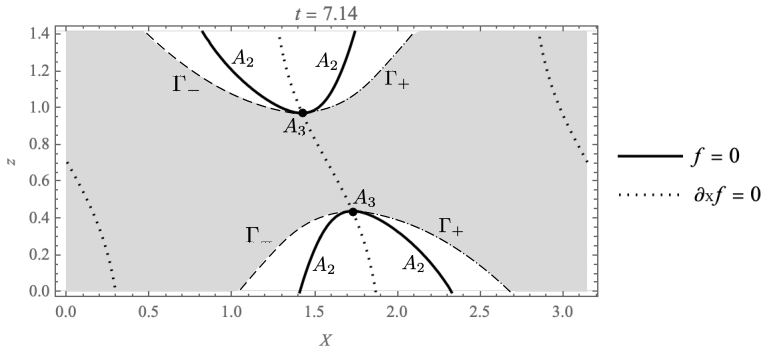}
    \caption{Region of physical interest in the $(x,z)$-plane for $t\approx 7.14$. The singular locus $\Sigma L$ and the boundaries $\gamma_-$ and $\Gamma_+$ are shown as a bold, dashed, and dash-dotted curve respectively. All the fold points $A_2$ are found outside the physical region, whereas the cusp points $A_3$ are found on its boundary.}
    \label{fig:physreg}
\end{figure}

As established above, the identification of the regularized Lagrangian submanifold $\cLt$ is a necessary step in the reconstruction of the velocity field. When working in coordinates, this amounts to finding the region of physical interest in the $(X,z)$-plane (see Figure \ref{fig:physreg}). The boundaries of the physical region $\Gamma_+,\Gamma_-$ are obtained by Chynoweth and Sewell's equations (\ref{X1 X2 system}). These are solved numerically for several values of $z$ and $t$ for solutions $X_+(z,t),X_-(z,t)$ within the observation window (\ref{period}). Figure \ref{fig:physreg} shows the region of physical interest at a particular time and its relationship with the singular set. It is worth noting that all the $A_2$ points of $\Sigma L_t$ fall outside the physical region (i.e., $\cLt$) while the $A_3$ points are found on its boundary (i.e., on $\partial\cLt$). This has implications on the velocity field, as we discuss next.

\begin{figure}
    \centering
    \includegraphics[width=0.4\textwidth]{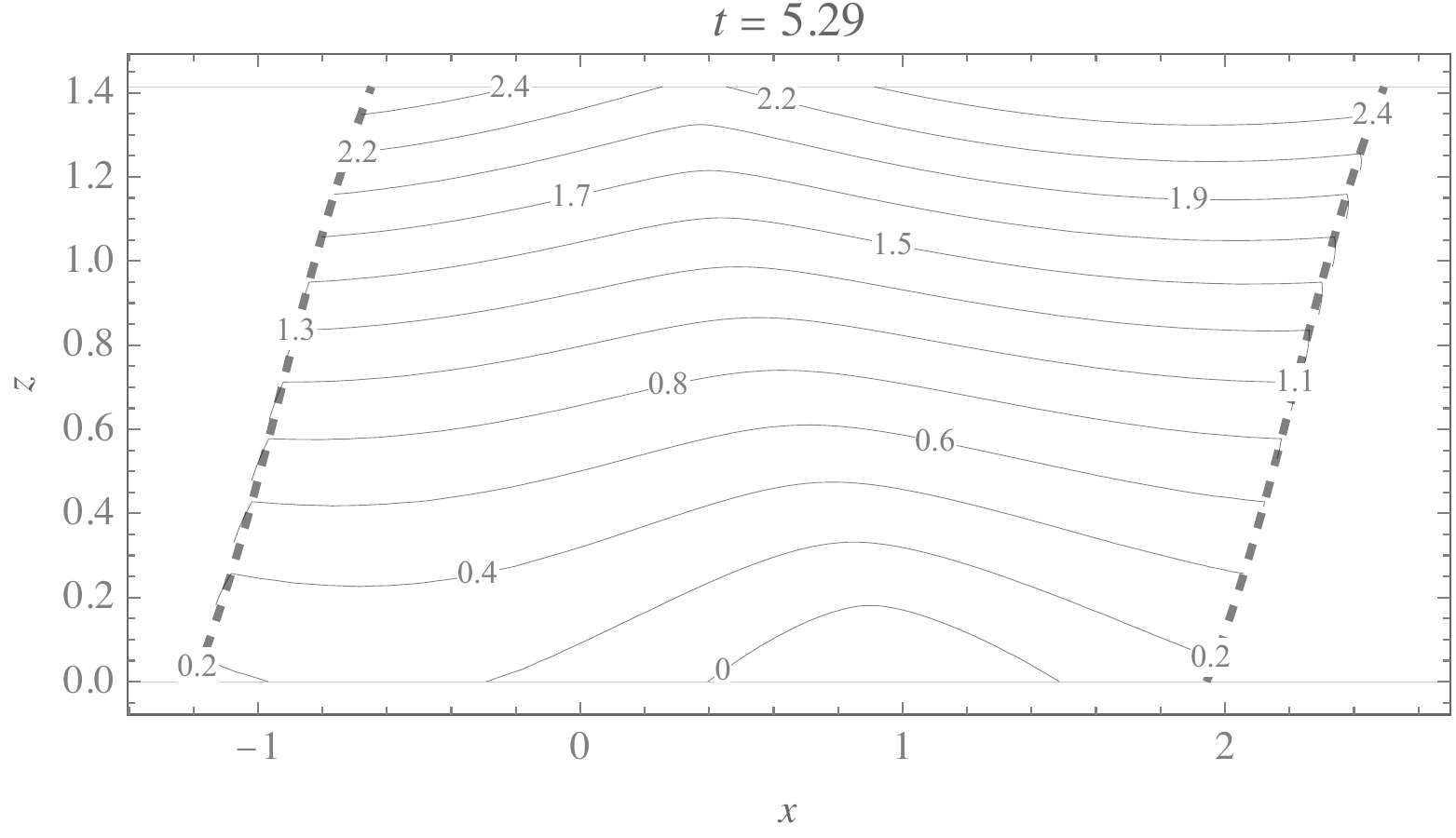}
    \includegraphics[width=0.4\textwidth]{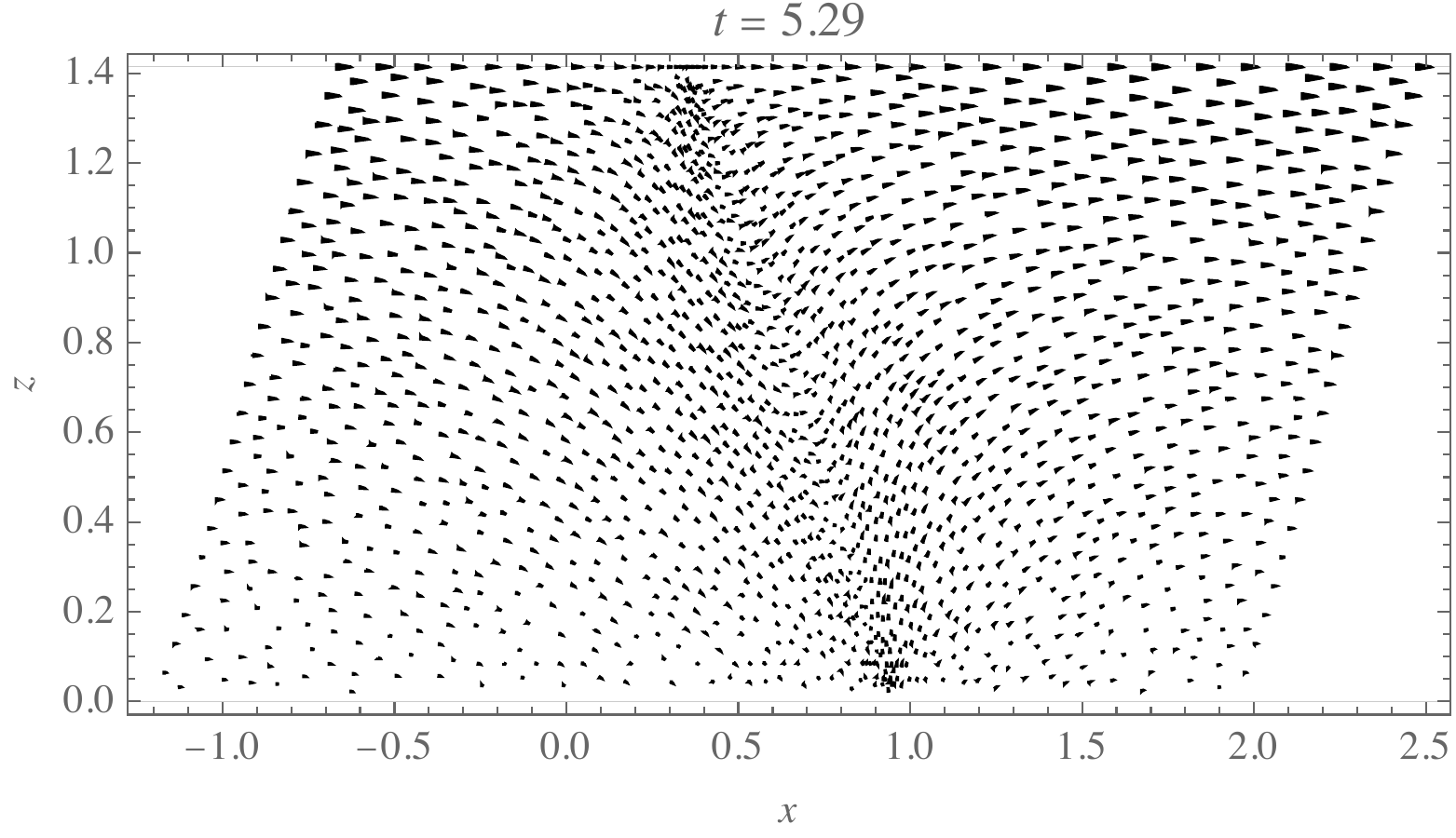}
    \includegraphics[width=0.4\textwidth]{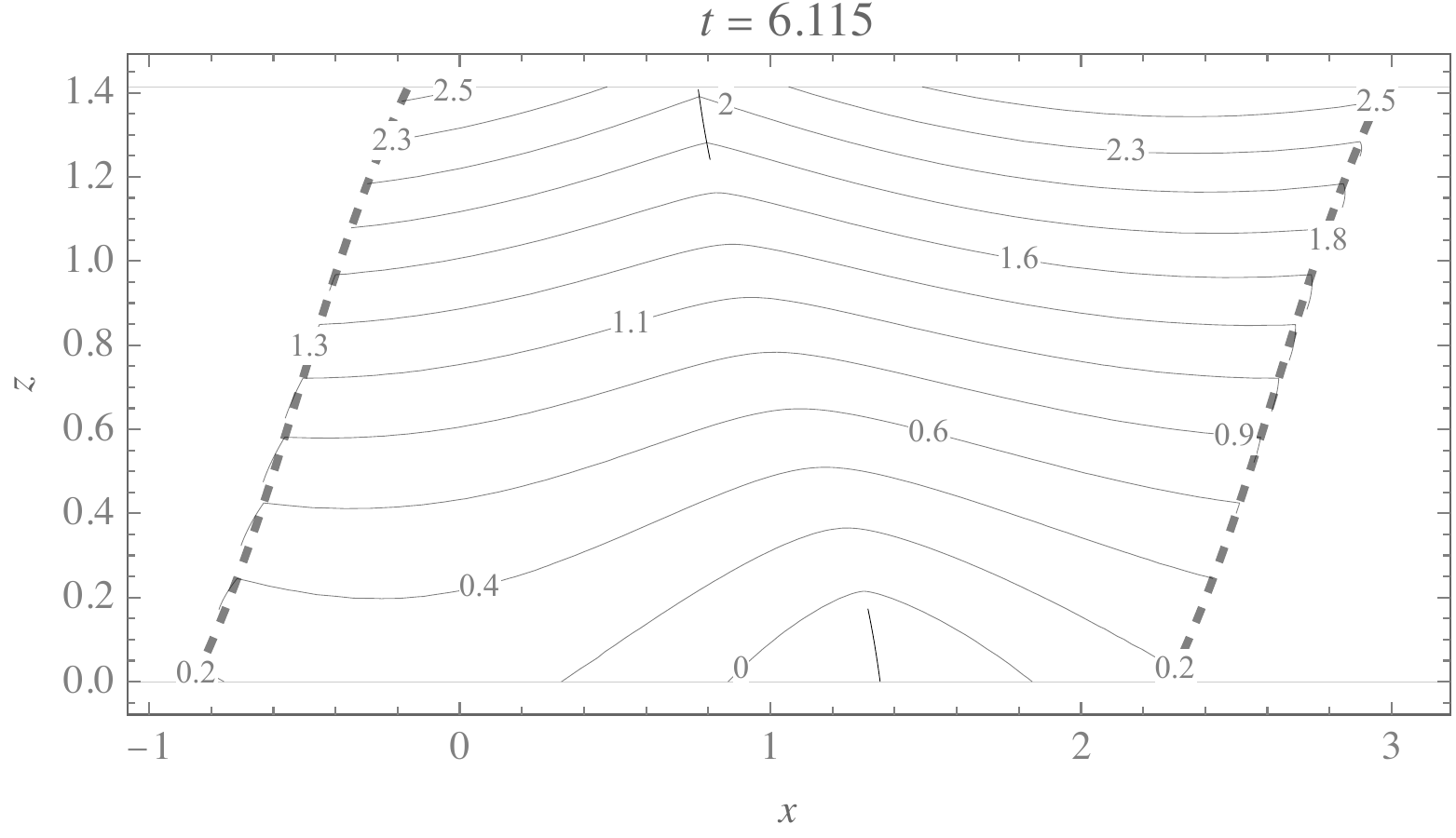}
    \includegraphics[width=0.4\textwidth]{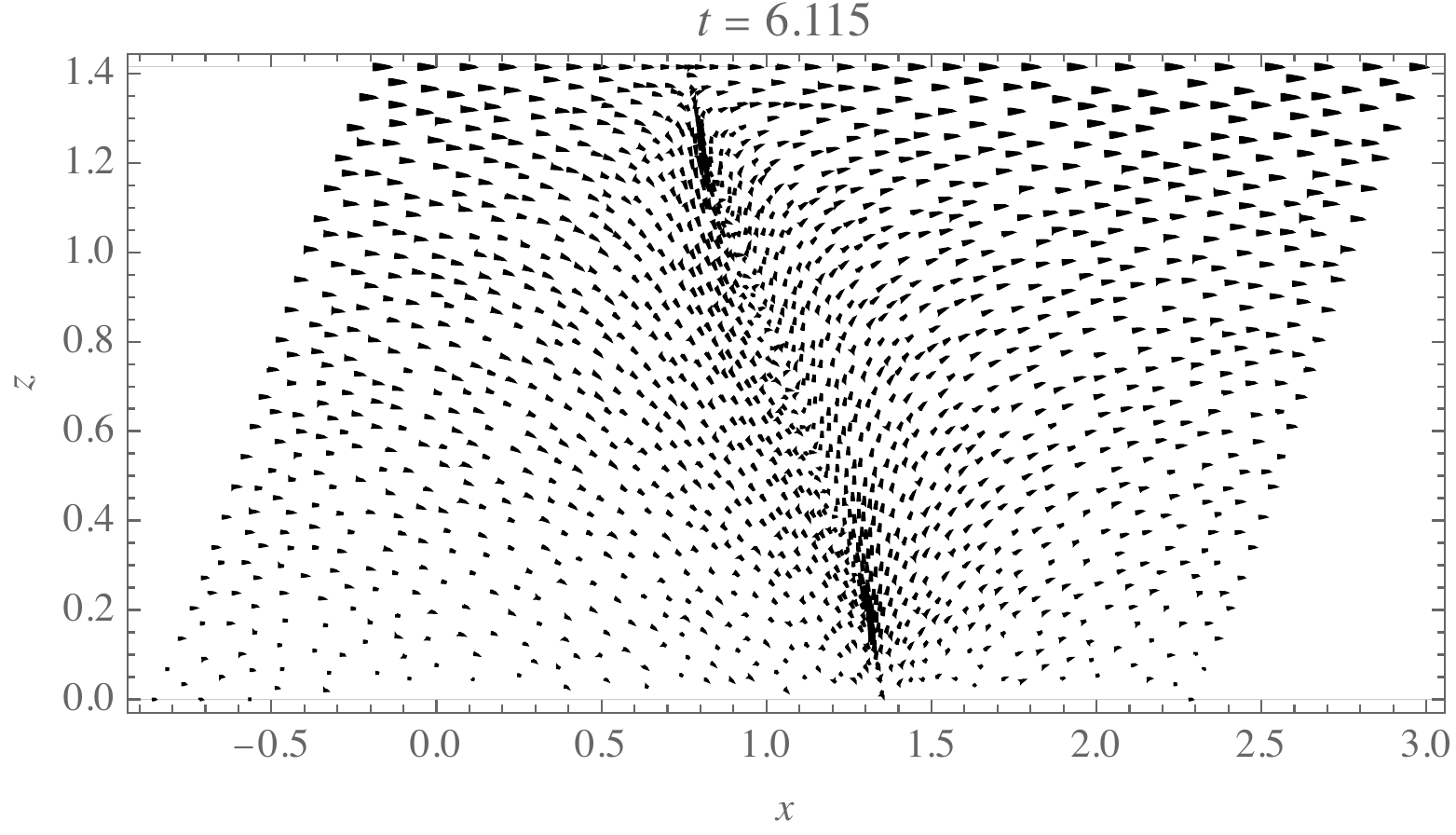}
    \includegraphics[width=0.4\textwidth]{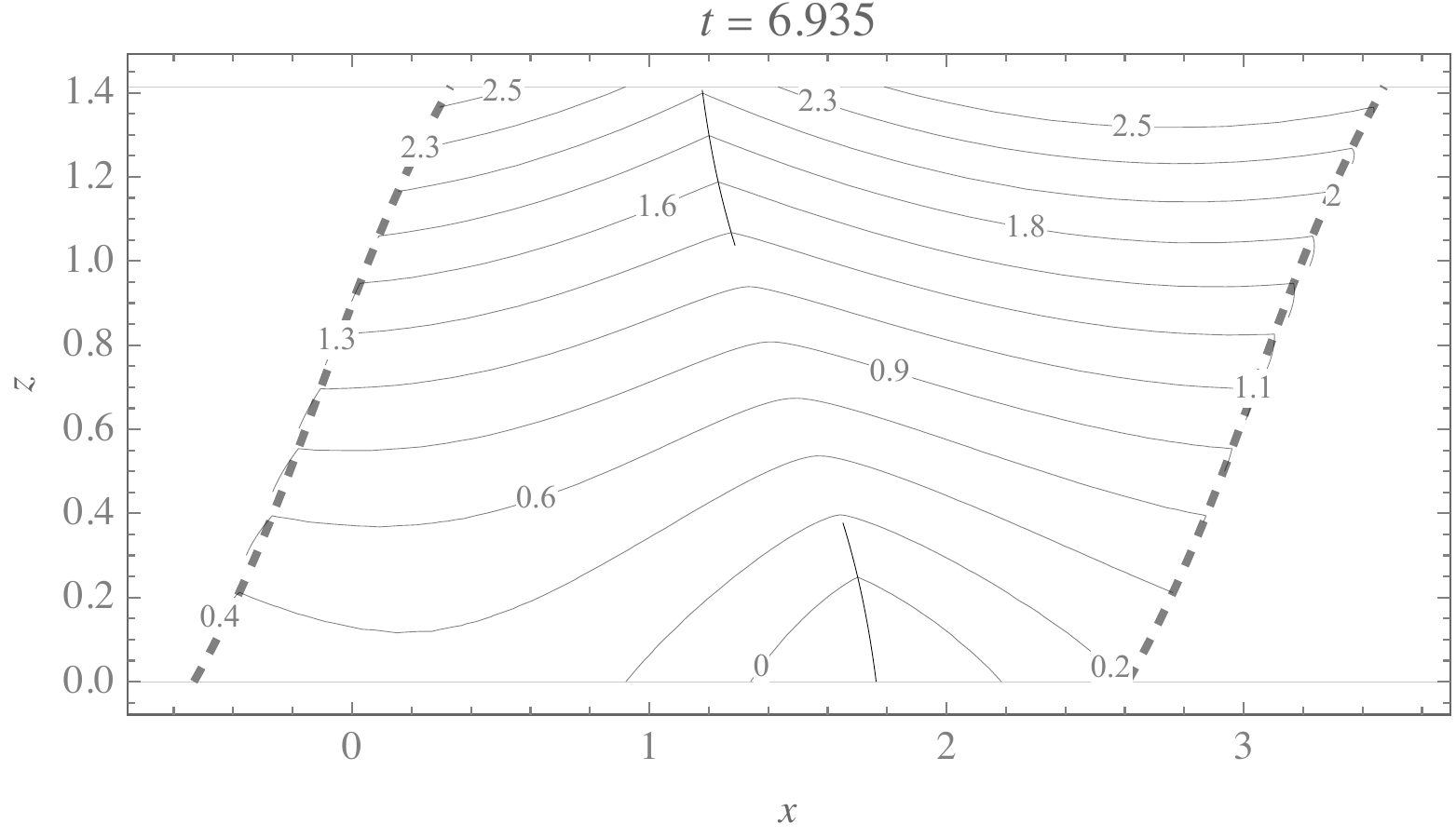}
    \includegraphics[width=0.4\textwidth]{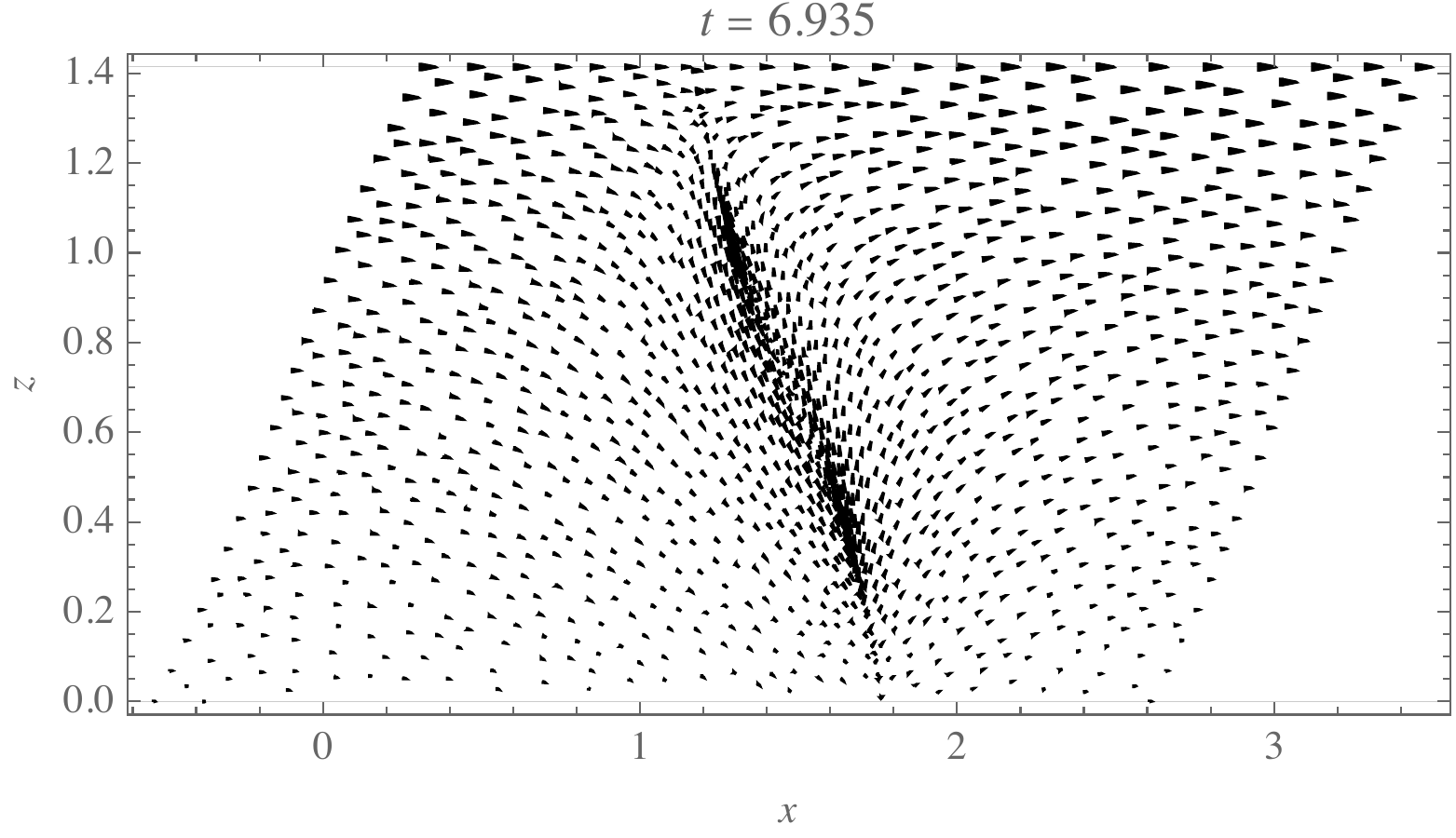}
    \includegraphics[width=0.4\textwidth]{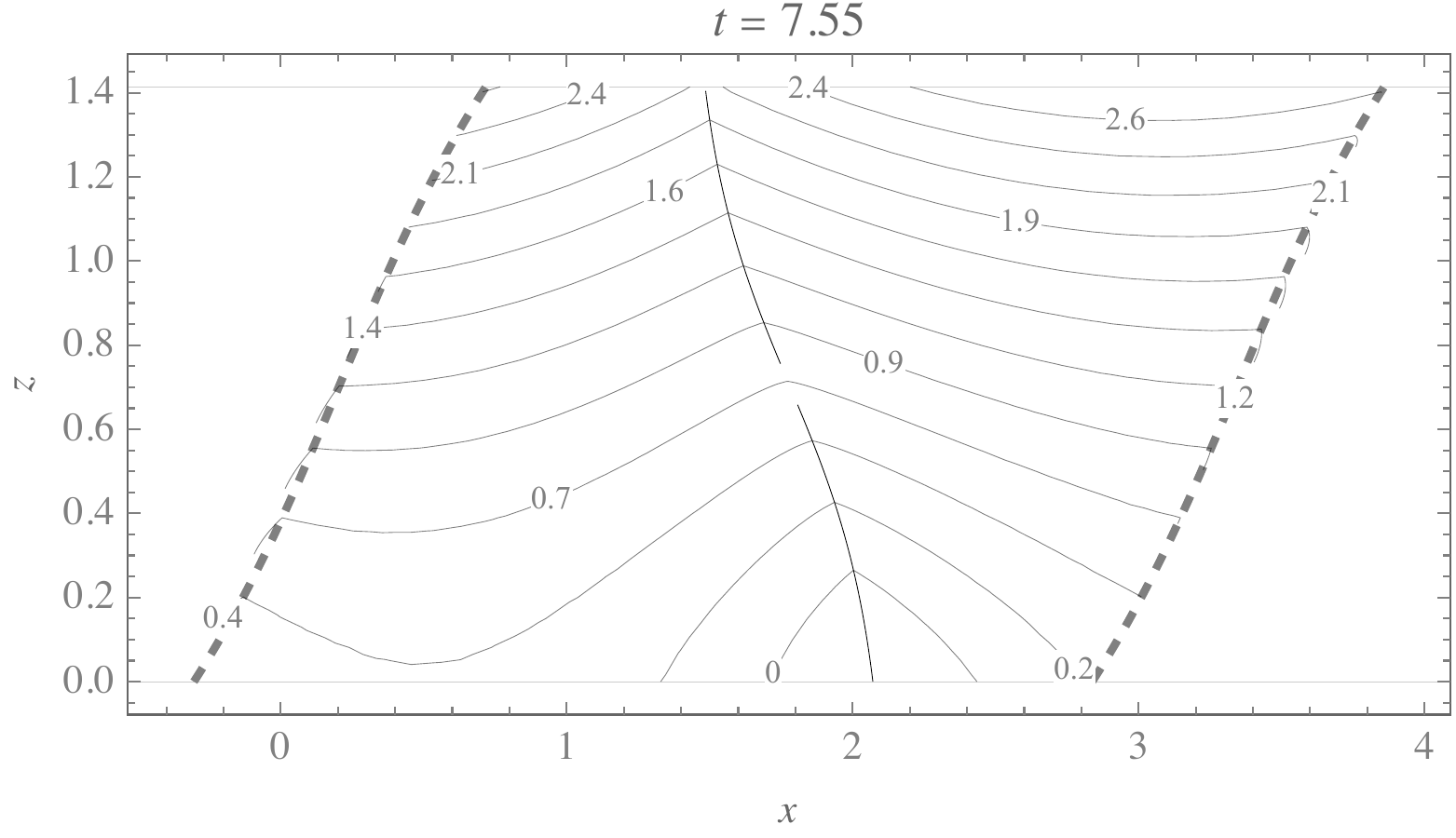}
    \includegraphics[width=0.4\textwidth]{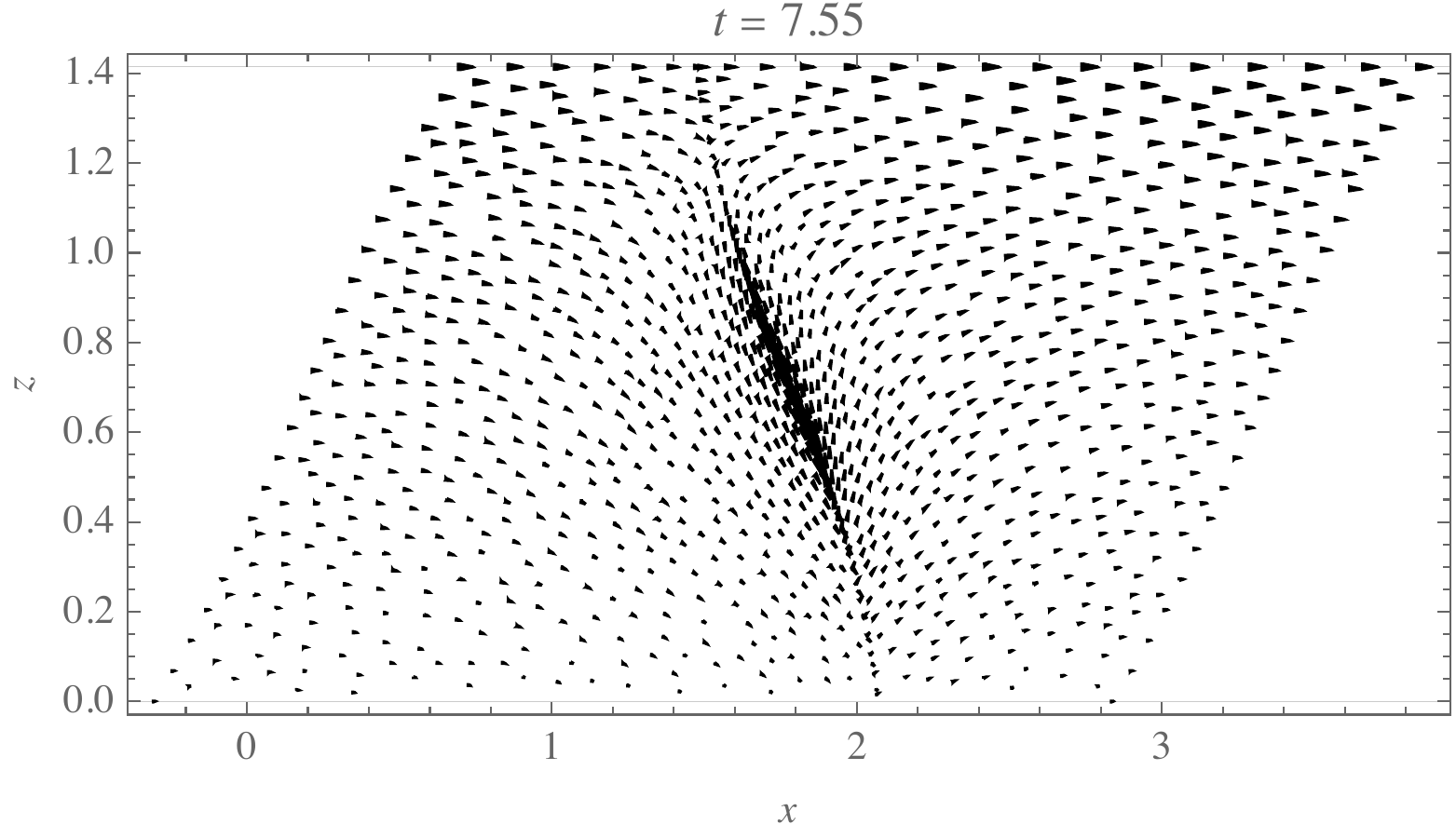}
    \includegraphics[width=0.4\textwidth]{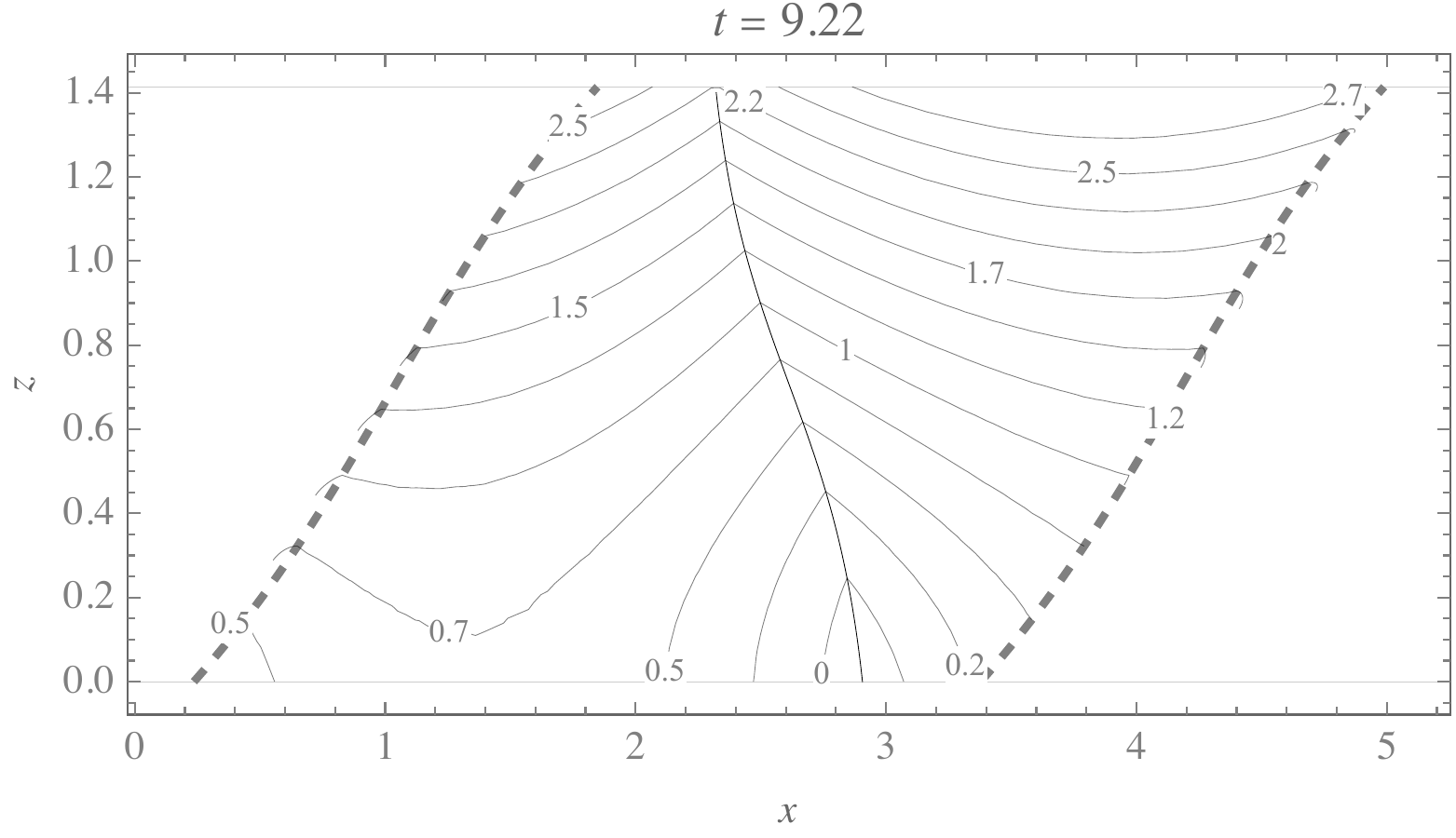}
    \includegraphics[width=0.4\textwidth]{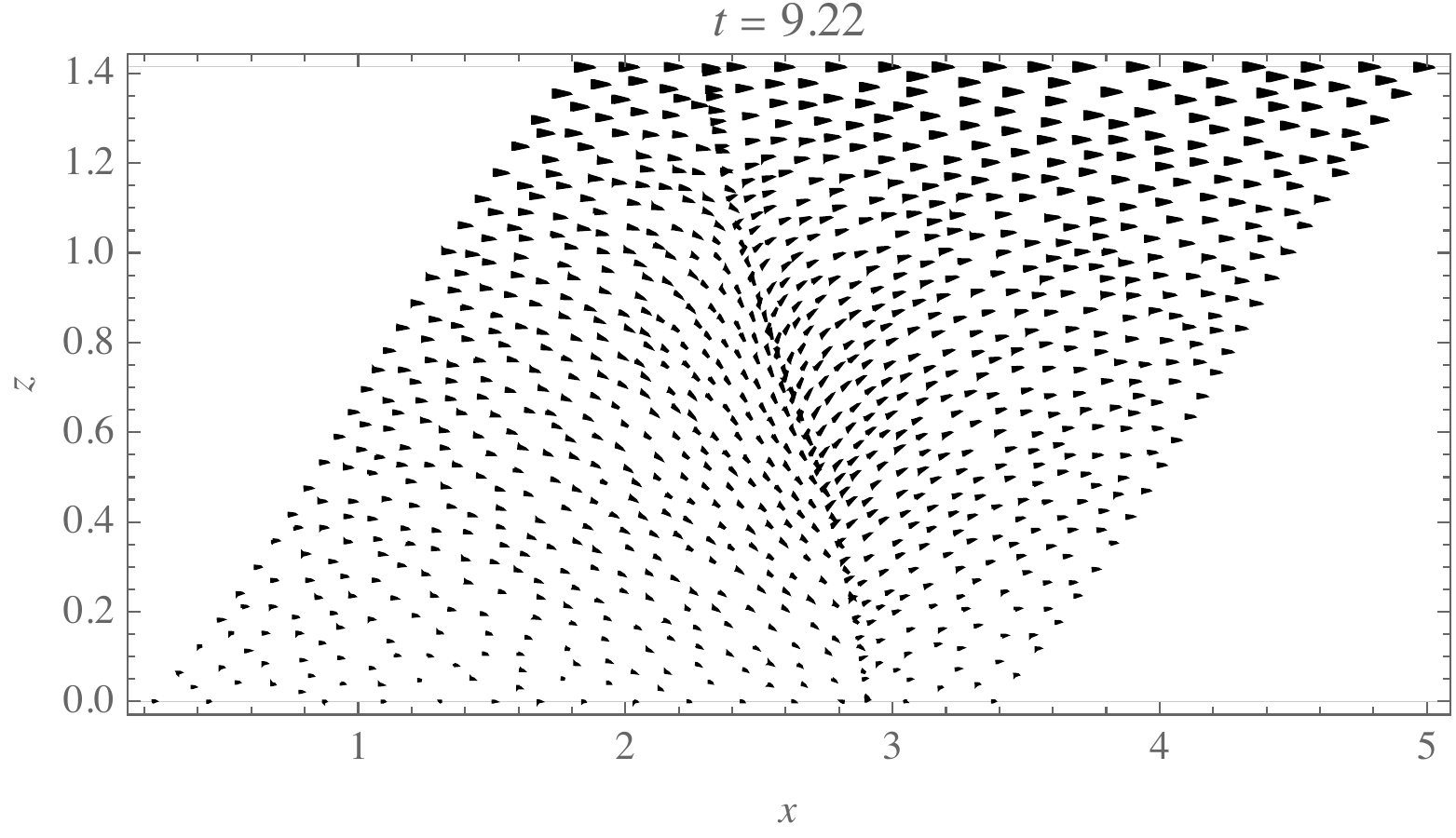}
    \caption{Left: contours of the geopotential $\phi$ on the plane $y=0$ for different times. The observation window is deformed by the projection mapping $\piL$ (i.e., the inverse Legendre transform) and its boundaries are indicated by dashed lines. \CS fronts are represented by thin black curves. Right: the corresponding velocity field $(u,w)$. The front corresponds to the region where the density of plotted points is higher. This is a by-product of the projection operation, and has no intrinsic physical meaning.}
    \label{fig:velocity}
\end{figure}

Figure \ref{fig:velocity} provides a view of the physical picture associated with $\cLt$. The physical velocity field $(u,w)$ on the slice $y=0$ is shown along with the corresponding geopotential $\phi$ for several times. 
At $t\approx 5.29<t'$, the solution is still regular -- the geopotential contours and the velocity field are smooth -- but the singularity is about to occur.
For $t\ge t'$, the solution contains frontal surfaces, where both the velocity field and the slope of the geopotential contours experience a jump discontinuity. Because (the closure of) $\cLt$ shares the $A_3$ points with $\Sigma L$, the velocity field blows up at the projection of these points, which represent the fronts tip.
The fronts keep growing over time until $t=t''$, when they meet and become a single front running continuously from $z=0$ to $z=\sqrt{2}$. The velocity field is everywhere bounded for $t>t''$.

\section{Pseudo-Riemannian geometry}
\label{sec:LR metric}

In a previous work \cite{DOR2023}, a new approach to the classification of singularities of semigeostrophic equations was presented. This point of view, based on pseudo-Riemannian geometry, was used to study a stationary solution containing a fold singularity. In this section, we apply this framework to the Eady problem, an example of a non-stationary solution which becomes singular over time.

\subsection{The \LR metric}
In order to classify symplectic \MA equations in three variables, Lychagin and Rubtsov \cite{LR83} introduced a pseudo-Riemannian metric on the phase space whose signature uniquely identifies the various classes. For a given \MA structure $(\alpha,\omega)$, the \LR metric $g_\effective$ can be defined through the equation
\begin{equation}
\label{LR metric 3d}
    g_\effective(\xi_1,\xi_2)\frac{\symplectic\wedge\symplectic\wedge\symplectic}{3!}=\iota_{\xi_1} \effective\wedge \iota_{\xi_2}\effective\wedge \symplectic,
\end{equation}
which holds for any pair of vector fields $\xi_1,\xi_2$ on the phase space. For the semigeostrophic \MA structure (\ref{MA structure}), the \LR metric on $\cotb{3}$ takes the simple form
\begin{equation}
\label{LR metric SG}
	g_\effective=2\PV(dxdX+dydY+dzdZ).
\end{equation}
The potential vorticity is a positive constant for the Eady solution, thus the $g_\effective$ is a flat metric with constant signature across $\cotb{3}$. A much more interesting object is the pull-back $h_\effective$ of $g_\effective$ onto the Lagrangian submanifold $L_t$ that represents an unstable Eady wave. The pull-back metric $h_\effective$ reads, in local coordinates $(X,Y,z)$ on $L_t$,
\begin{equation}
\label{LR metric S}
	h_\effective=2\PV \bigg(\frac{\partial^2S}{\partial X^2}dX^2+2\frac{\partial^2S}{\partial X\partial Y}dXdY+\frac{\partial^2S}{\partial Y^2}dY^2-\frac{\partial^2S}{\partial z^2}dz^2\bigg),
\end{equation}
where $S(X,Y,z,t)$ is the generating function for $L_t$. As discussed in \cite{DOR2023}, there are a few properties of $h_\effective$ which hold for all the generalized solutions to the semigeostrophic equations. In a nutshell, $h_\effective$ is Riemannian on elliptic branches of $L_t$ and pseudo-Riemannian on hyperbolic ones. Moreover, $h_\effective$ degenerates on parabolic branches, i.e., the singular set $\Sigma L_t$.



\subsection{Curvature}

The potential vorticity $\PV$ is constant for Eady solutions. Equation (\ref{LR metric SG}) then implies that the ambient space is flat. However, the Lagrangian submanifolds which represent these solutions can in general be curved. It turns out that the sign of the scalar curvature of an Eady solution $L_t$ is directly connected to the signature of the pull-back metric, a feature that reveals the essential 2-dimensional nature of these solutions. We prove this statement with regards to $X$-travelling waves, and then extend our conclusions to general Eady waves by means of an adapted coordinate system.

\begin{figure}
\center
\includegraphics[width=0.24\textwidth]{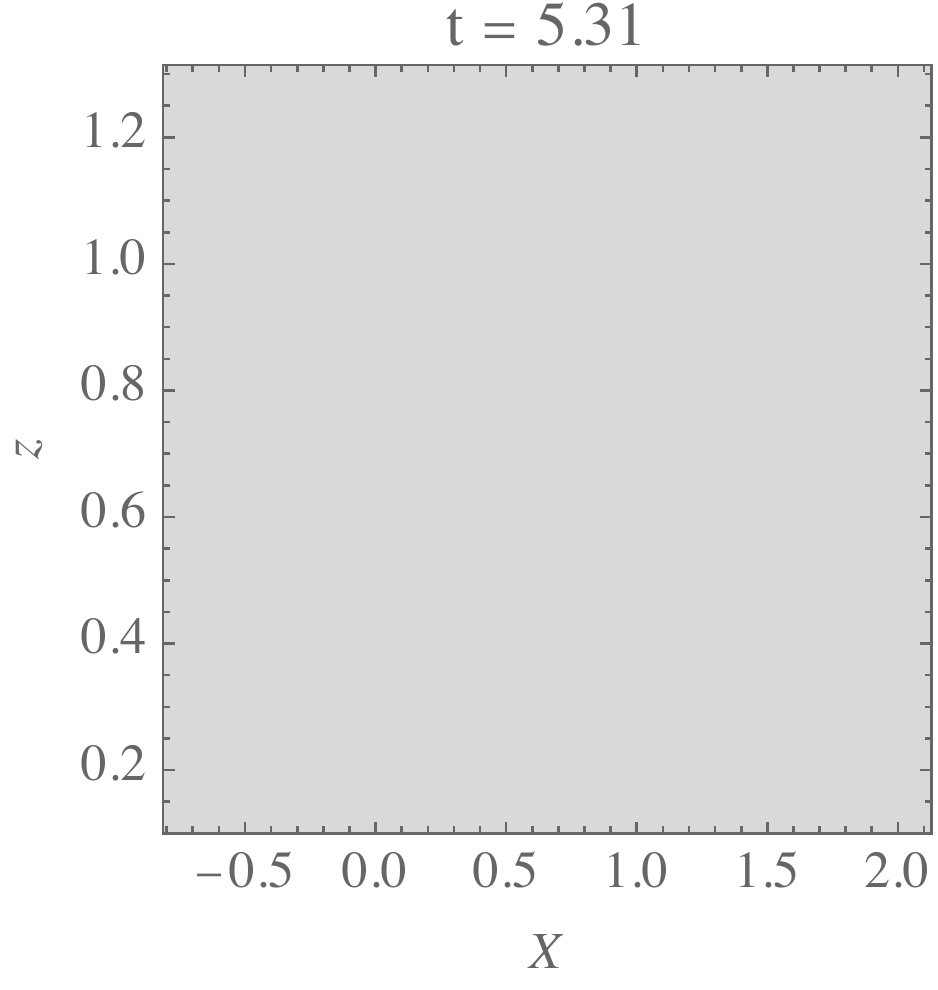}
\includegraphics[width=0.24\textwidth]{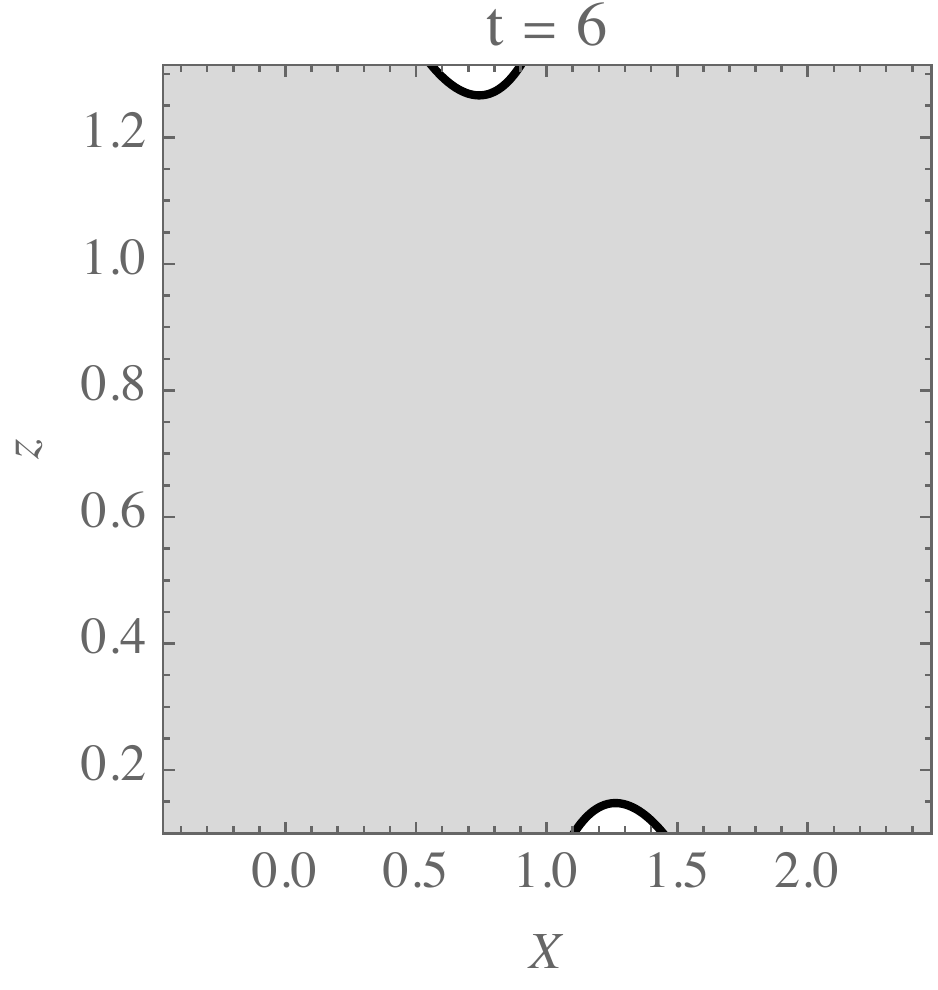}
\includegraphics[width=0.24\textwidth]{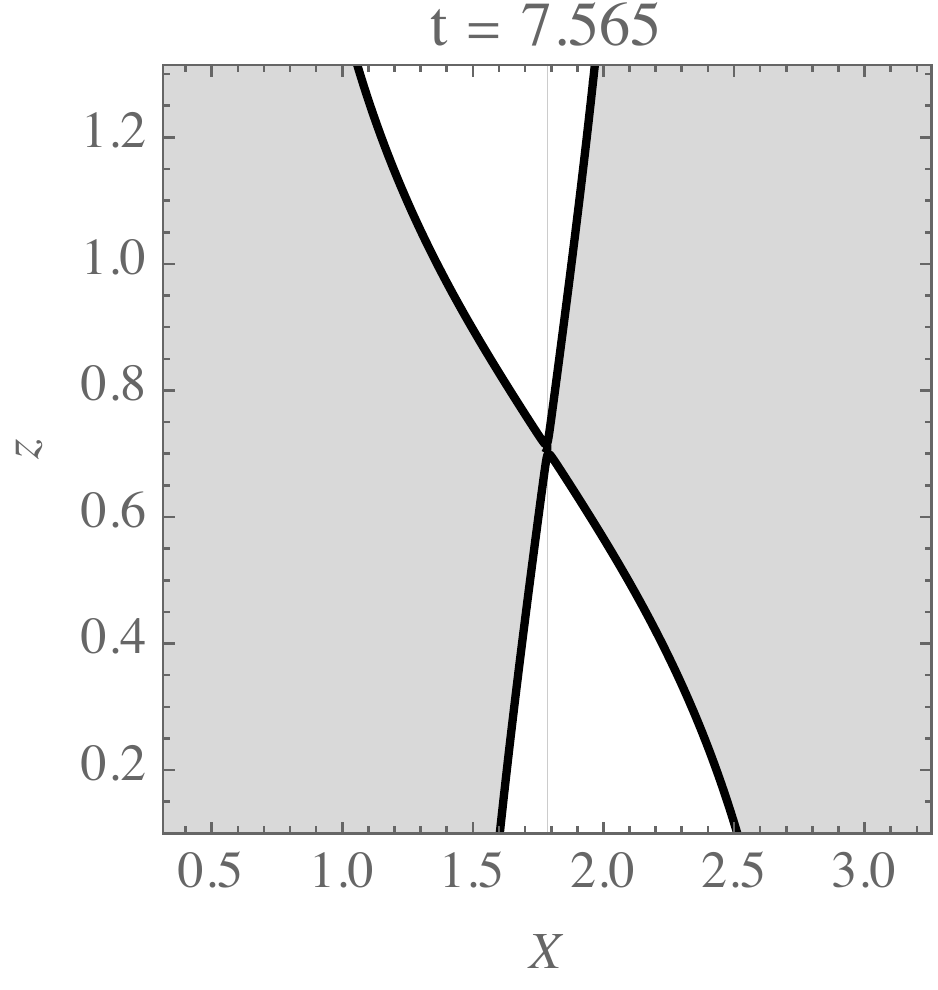}
\includegraphics[width=0.24\textwidth]{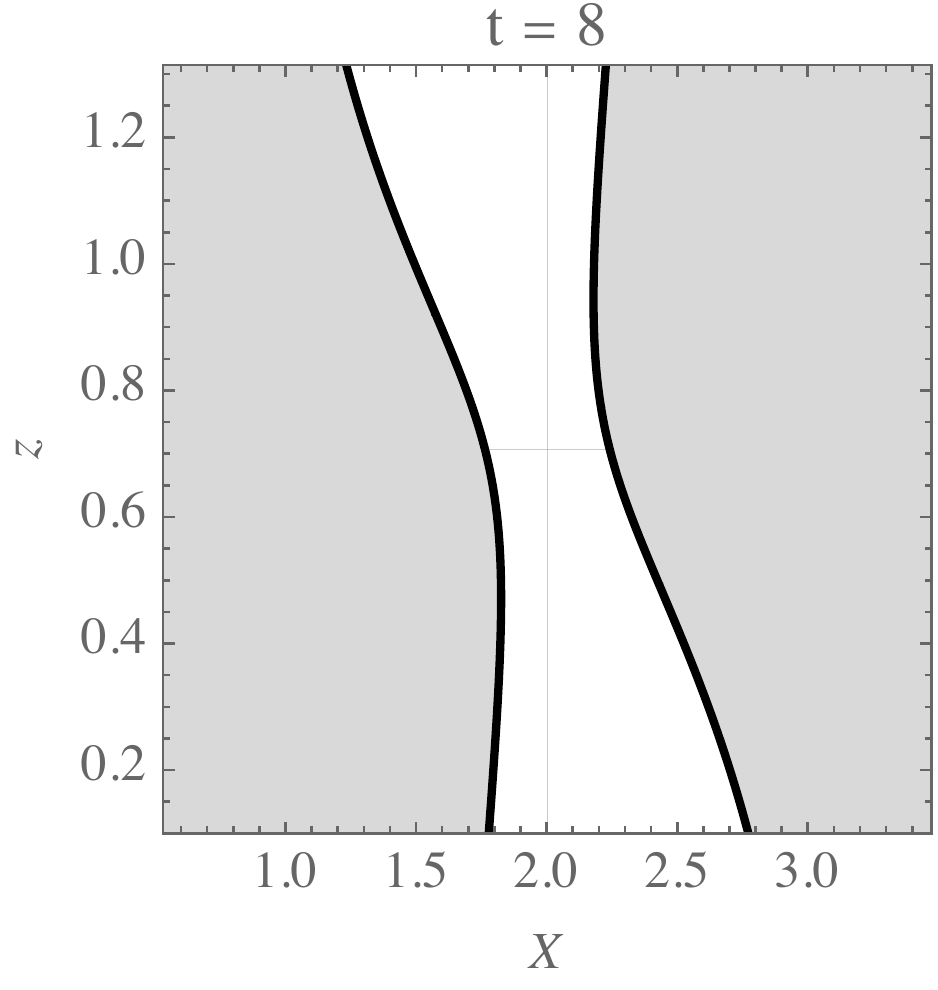}
\caption{The regions of positive scalar curvature (gray) within a constant-$Y$ slice of $L_t$ at different times. The black boundaries represent (a slice through) the singular locus $\Sigma L_t$.}
\label{Fig:sign Sc}
\end{figure}

\begin{figure}
\includegraphics[width=0.3\textwidth]{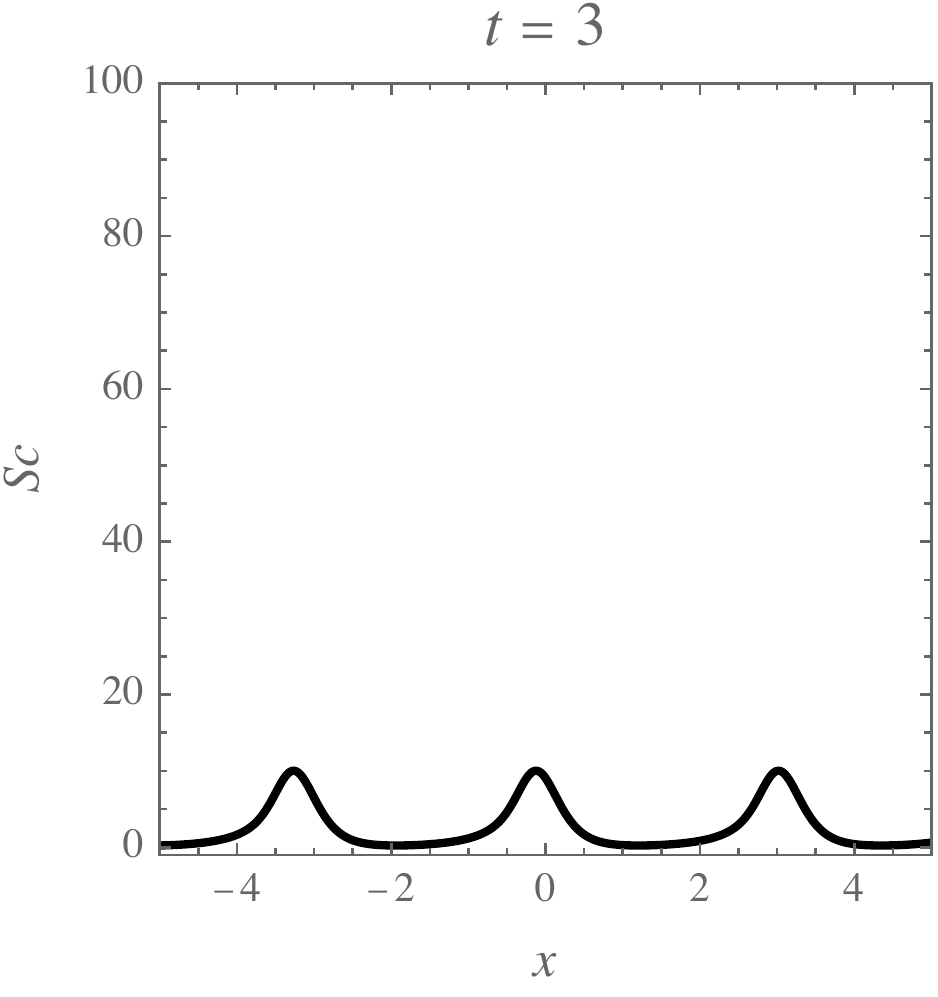}
\hspace{10pt}
\includegraphics[width=0.3\textwidth]{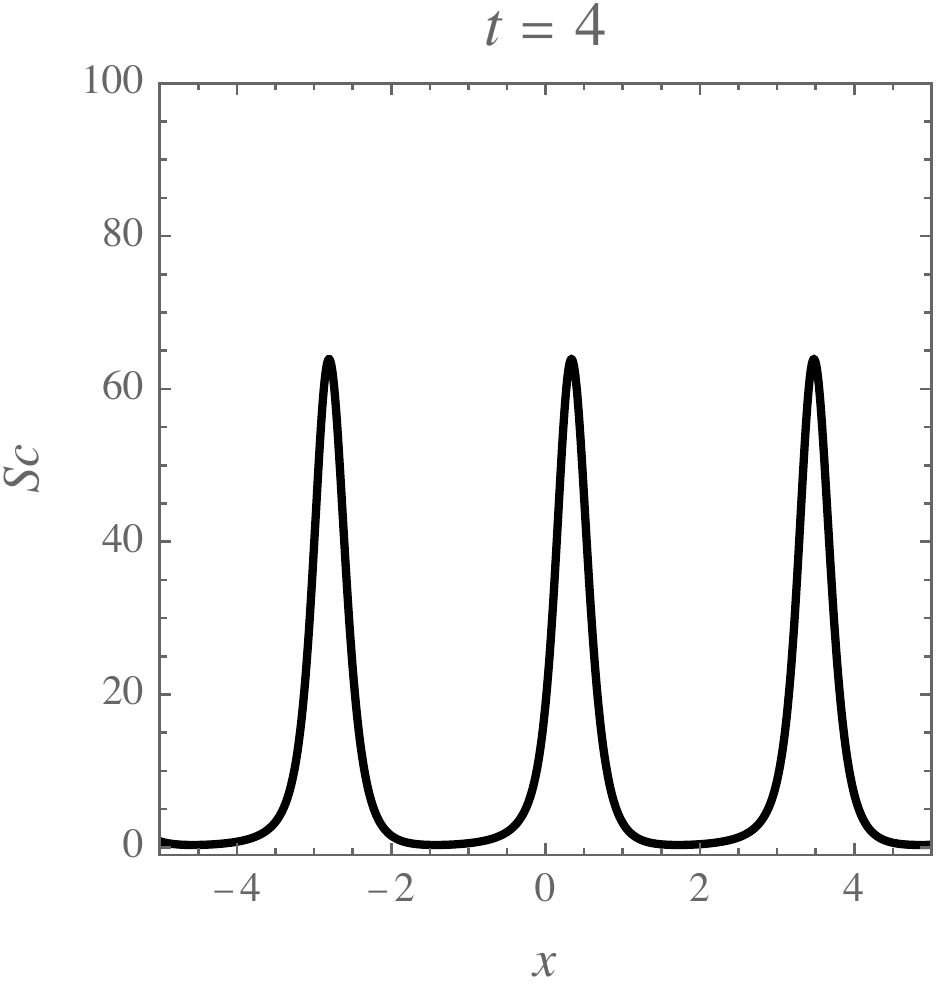}
\hspace{10pt}
\includegraphics[width=0.3\textwidth]{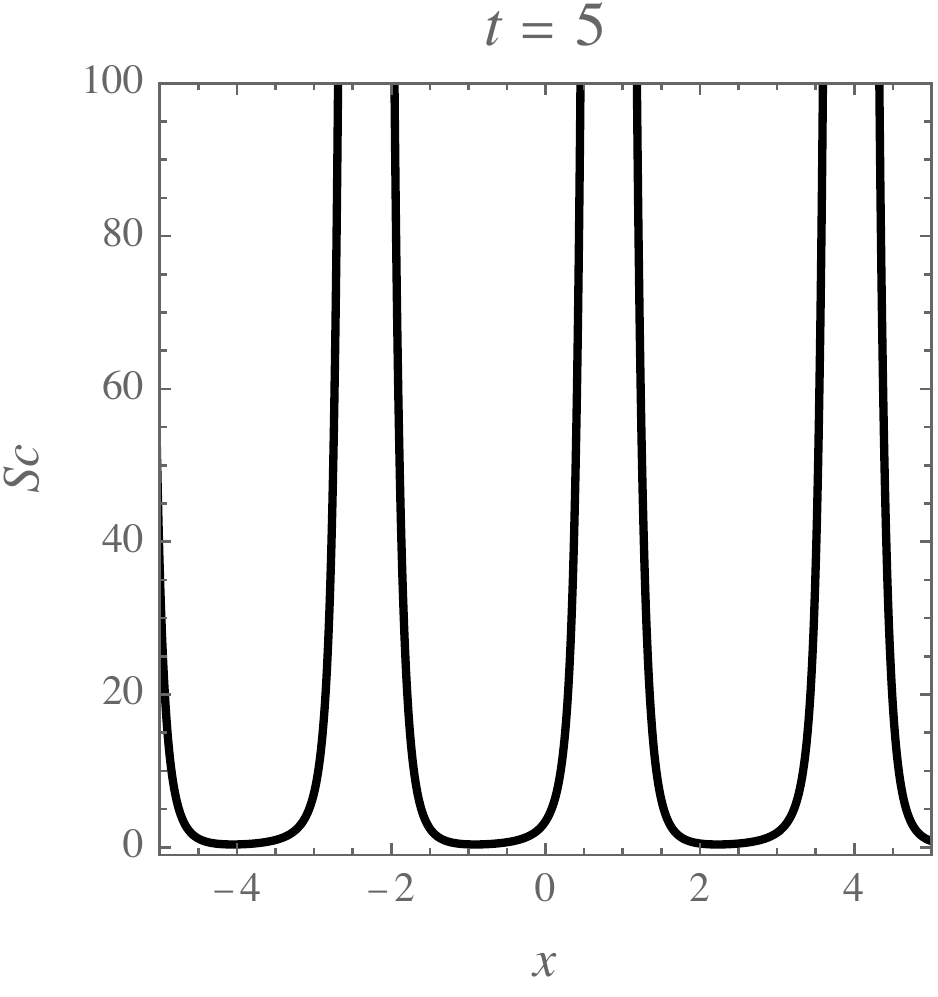}
\label{Fig:Sc vs X}
\caption{Slice at $z=0$ through the graph of $\scalar$ for three times preceding $t'\approx 5.3$. Curvature maxima are already visible at $t=3$ and grow unbounded quickly as $t\to t'$.}

\end{figure}

Waves travelling in the $X$-direction are specified by a generating function of the form
\begin{equation}
\label{implicit X-waves}
    S=\frac{Y^2}{2}+\fr\,Yz+S'(X,z,t),
\end{equation}
where $S_0$ is the Eady basic state (\ref{S0}), and $S_1$ is the perturbation field (\ref{ansatz S1}). 
The pull-back metric (\ref{LR metric S}) thus reads
\begin{equation}
    h_\effective=2\PV\bigg(\frac{\partial^2S'}{\partial X^2}dX^2+dY^2-\frac{\partial^2S'}{\partial z^2}dz^2\bigg),
\end{equation}
and, since $S'$ satisfies (\ref{CS S 2d}), it may be simplified to
\begin{equation}
\label{h X-waves}
    h_\effective=2\PV\bigg(\frac{\partial^2S}{\partial X^2}(dX^2+\PV dz^2)+dY^2\bigg).
\end{equation}
Because $\PV$ is a constant, it follows from (\ref{h X-waves}) that the geometry of a Lagrangian submanifold $L_t$ representing an $X$-wave locally decomposes into the Cartesian product of the $Y$-axis and a slice $Y=const.$
The scalar curvature of the constant-$Y$ slices (which coincides with the scalar curvature of the whole $L_t$) is\footnote{Curvature calculations were performed using Professor Leonard Parker’s Mathematica notebook “Curvature and the Einstein Equation,” available online \href{https://web.physics.ucsb.edu/~gravitybook/mathematica.html}{here} \cite{Parker}.}
\begin{gather}
\label{scalar}
    \scalar=\frac{\big(\frac{\partial f}{\partial z}\big)^2+\PV\big(\frac{\partial f}{\partial X}\big)^2-f\big(\frac{\partial^2f}{\partial z^2}+\PV\frac{\partial^2f}{\partial X^2}\big)}{\PV f^3},
\end{gather}
where $f\equiv\partial_X^2S$ is defined in (\ref{def f}), and, with the choice of coefficients (\ref{coeff choice}), reads
\begin{equation}
    f=1-8\sqrt{2}e^{\omega_i t}\bigg[2e^{-\sqrt{2}z}\cos(t-2X)+\omega_i(e^{\sqrt{2}z}+e^{-\sqrt{2}z})\sin(t-2X)\bigg]\eta.
\end{equation}
Now observe that the second addendum in the numerator of (\ref{scalar}) is identically zero. Indeed,
\begin{equation}
    \frac{\partial^2f}{\partial z^2}+\PV\frac{\partial^2f}{\partial X^2}=\frac{\partial^3S'}{\partial X\partial z^2}+\PV\frac{\partial^3S'}{\partial X^3}=\frac{\partial}{\partial X}\bigg(\frac{\partial^2 S'}{\partial z^2}+\PV\frac{\partial^2 S'}{\partial X^2}\bigg)=0,
\end{equation}
where we have used $\PV=constant$ and (\ref{CS S 2d}).
We are thus allowed to write (\ref{scalar}) in the form
\begin{gather}
\label{scalar 2}
    \scalar=\frac{\left(\frac{\partial f}{\partial z}\right)^2+\PV \left(\frac{\partial f}{\partial X}\right)^2}{\PV f^3}.
\end{gather}
Formula (\ref{scalar 2}) brings up some basic facts about the curvature of $X$-travelling waves:
\begin{itemize}
	\item
	Using (\ref{CS S}) and the last of (\ref{grad S}) in (\ref{def f}), we may write $f$ as
	\begin{equation}
 \label{strat to vort ratio}
		f=\frac{\partial\theta/\partial z}{\PV}=\frac{N^2}{\PV}.
	\end{equation}
    The last equality follows from the definition of the Brunt--Vaisala buoyancy frequency (see for example \cite{Hos75}).
    The numerator in this expression represents a measure of the atmosphere stratification, and is directly related to the static stability of the atmosphere ($N^2>0$ for a stably stratified atmosphere). For constant $\PV>0$ and $N^2>0$, (\ref{scalar 2}) may be seen as a norm of the gradient $(\partial_XN,\partial_zN)$.
    An alternative way to look at (\ref{scalar 2}) is by assigning $\scalar$ and looking for a function $N^2$ that verifies the equality. In this case, (\ref{scalar 2}) takes the form of a Hamilton--Jacobi equation for $N^2$.
	\item As the numerator in (\ref{scalar 2}) is always non-negative, the sign of $\scalar$ is inherited from $f$. This puts $\scalar$ in relation with the signature of $h_\effective$: the Riemannian branches of $L$ are positively curved, whereas the psuedo-Riemannian branches of $L$ are negatively curved (see Figure \ref{Fig:sign Sc}). This can also be stated in terms of ellipticity/hyperbolicity of the \MA operator: the elliptic branches of $L$ are positively curved whereas the hyperbolic branches of $L$ are negatively curved \cite{DOR2023}.
	\item The numerator in (\ref{scalar 2}) is a strictly positive quantity at generic points on $\Sigma L=\{f=0\}$, which implies that $\scalar$ blows up at generic singular points (including $A_3$ points). This statement fails at $t=t''$, when the topology of the singular locus changes. At such time, higher order singularities are produced (see Figure \ref{Fig:sign Sc}) and the numerator in (\ref{scalar 2}) vanishes. It can be verified that the numerator in (\ref{scalar 2}) goes to zero faster than $f^3$ as one of these points is approached, which means that $\scalar\to 0$ as well.
	\item The scalar curvature blows up at $A_3$ points (as $f^{-3}$). Thanks to this, $\scalar$ may be interpreted as a diagnostic tool which reveals the impending development of a front. Figure \ref{Fig:Sc vs X} shows a section through the graph of $\scalar(X,z,t)$ for fixed $z$ at different times before the catastrophe time. The maxima appear much earlier than the swallowtail singularities do, and grow unbounded quickly. Tracing them can give an estimate of the position where the fronts will appear.
    \item As formula (\ref{scalar 2}) indicates, the scalar curvature is a measure of the gradient of $N^2$ ($\PV$ is a constant here). This implies that $\scalar$ vanishes at critical points of $N^2$.
    This statement may be compared with the results of Napper et al. \cite{NRR2023}, where the sign of the scalar curvature was linked to the maxima and minima of the Laplacian of pressure in the 2D Euler equations.
    Further work is needed to investigate the relationship between $\scalar$ and local accumulation of buoyancy $N^2$.
\end{itemize}

We conclude this section by extending these conclusions to solutions that represent Eady waves travelling in a generic direction (except $Y$). If $l\ne 0$, then $S_1$ depends also on $Y$, and neither (\ref{h X-waves}) nor (\ref{scalar 2}) are valid anymore. However, it is always possible to introduce a rotated coordinate system on $L_t$ which allows us to write 
the pull-back metric and the scalar curvature in a form analogous to (\ref{h X-waves}) and (\ref{scalar 2}). 
Namely, we introduce a change of coordinates on $\cotb{3}$,
\begin{equation}
    (x,y,z,X,Y,Z)\mapsto (x,y,z,X',Y',Z),
\end{equation}
by
\begin{equation}
\label{rot coords}
    X'=\frac{k X+l Y}{m},\qquad Y'=\frac{k Y-l X}{m},
\end{equation}
with inverse
\begin{equation}
    X=\frac{kX'-lY'}{m},\qquad Y=\frac{lX'+kY'}{m}.
\end{equation}
Observe that
\begin{equation}
    dX\wedge dY=\bigg(\frac{k}{m}dX'-\frac{l}{m}dY'\bigg)\wedge\bigg(\frac{l}{m}dX'+\frac{k}{m}dY'\bigg)=
\end{equation}
\begin{equation}
    =\frac{k^2}{m^2}dX'\wedge dY'-\frac{l^2}{m^2}dY'\wedge dX'=\frac{k^2+l^2}{m^2}dX'\wedge dY'=dX'\wedge dY'.
\end{equation}
Therefore, the \MA form $\alpha$ (\ref{MA structure}) becomes
\begin{equation}
    \effective=dX'\wedge dY'\wedge dZ-\PV dx\wedge dy\wedge dz.
\end{equation}
The first two equations defining $L_t$ (see (\ref{Lagr S})) become
\begin{equation}
\label{L syst rot}
    x=\frac{\partial S}{\partial X}=\frac{k}{m}\frac{\partial S}{\partial X'}-\frac{l}{m}\frac{\partial S}{\partial Y'},\qquad y=\frac{\partial S}{\partial Y}=\frac{l}{m}\frac{\partial S}{\partial X'}+\frac{k}{m}\frac{\partial S}{\partial Y'},
\end{equation}
with the last one remaining the same,
\begin{equation}
    Z=-\frac{\partial S}{\partial z}.
\end{equation}
Using these equations in $\alpha|_{L_t}=0$ yields
\begin{equation}
    \PV\bigg(\frac{\partial^2S}{\partial {X'}^2}\frac{\partial^2S}{\partial {Y'}^2}-\bigg(\frac{\partial^2S}{\partial X'\partial Y'}\bigg)^2\bigg)+\frac{\partial^2S}{\partial z^2}=0.
\end{equation}
This result says that the \CS equation (\ref{CS S}) is invariant under rotations of the $(X,Y)$-plane.
Next, consider the \LR metric (\ref{LR metric SG}). 
The ambient metric becomes
\begin{equation}
    \frac{g_\effective}{2\PV}=dx\bigg(\frac{k}{m}dX'-\frac{l}{m}dY'\bigg)+dy\bigg(\frac{l}{m}dX'+\frac{k}{m}dY'\bigg)+dz dZ.
\end{equation}
Using (\ref{L syst rot}), a lengthy but straightforward calculation shows that the pull-back metric $h_\effective=g_\effective|_{L_t}$ becomes
\begin{equation}
    h_\effective=2\PV\bigg(\frac{\partial^2S}{\partial {X'}^2}d{X'}^2+2\frac{\partial^2S}{\partial X'\partial Y'}dX' dY'+\frac{\partial^2S}{\partial {Y'}^2}d{Y'}^2-\frac{\partial^2S}{\partial z^2}dz^2\bigg),
\end{equation}
which, again, retain the same form as (\ref{LR metric S}).
Next, note that the general Eady solution may be written in the new coordinates as
\begin{equation}
    S=S_0+\varepsilon\Re(S_1),
\end{equation}
where,
\begin{equation}
    S_0=\frac{{X'}^2}{2}+\frac{{Y'}^2}{2}-\PV \frac{z^2}{2}-z+\fr\,z(Y'\cos(\nu)+X'\sin(\nu)),
\end{equation}
\begin{equation}
    S_1=\psi(\tilde z)e^{i(m\tilde X'-\omega t)},
\end{equation}
and
\begin{equation}
    \tilde X':=X'-\frac{k}{m}\frac{\ro}{2}t.
\end{equation}
Therefore, the form of a general Eady solution only differs from that of a $X$-travelling wave by the addition of the term
\begin{equation}
    \fr\,X'z\sin(\nu)
\end{equation}
in $S_0$.
Accordingly, the pull-back metric on $L_t$ reads
\begin{equation}
\label{h gen dir}
    h_\effective=2\PV\bigg(\frac{\partial^2S}{\partial {X'}^2}(d{X'}^2+\PV dz^2)+d {Y'}^2\bigg),
\end{equation}
and the scalar curvature becomes
\begin{equation}
\label{Sc gen dir}
    \scalar=\frac{\big(\frac{\partial f'}{\partial z}\big)^2+\PV\big(\frac{\partial {f'}}{\partial X'}\big)^2}{\PV {f'}^3},\qquad f':=\frac{\partial^2S}{\partial{X'}^2}.
\end{equation}
Equation (\ref{h gen dir}) says that the geometry of $L_t$ is naturally decomposed into the Cartesian product of constant-$Y'$ slices and the $Y'$-axis. Since the curvature (\ref{Sc gen dir}) has the same form as (\ref{scalar 2}), all the considerations about the $X$-travelling waves readily apply to general Eady waves as well.

\section{Conclusions and future directions}

In this work we have presented the classical Eady problem using the modern language of \MA geometry, already explored in the semigeostrophic context in \cite{DOR2023}.
The motivation behind this work is the dynamic nature of the singularity involved, in contrast to the example considered in \cite{DOR2023}, where a stationary solution containing a singularity was studied.
We have formalized the concept of Chynoweth–Sewell front in terms of Monge—Amp\'ere geometry and examined the Eady wave solutions by endowing them with a pseudo-Riemannian metric in this geometric framework.
In this regard, we have shown that the scalar curvature of Eady waves (understood as Lagrangian submanifolds of phase space) is exclusively dependent on the signature of the pull-back metric on them. 
This reflects the cylindrical nature of such solutions implied by their plane wave features.

As remarked in (\ref{subsec:CS fronts}), the class of solution (\ref{solution class}), which includes the Eady waves, satisfies the 2D \CS equation (\ref{CS S 2d}) which, in the case of a constant $\PV$, is equivalent to the classical Laplace equation in two space dimensions. Thanks to this observation, we can assert that an entire class of solutions is obtainable by complementing a harmonic function $S'(X,z)$ by the monomial terms in (\ref{solution class}).
Any function $S(X,Y,z)$ so obtained represents a solution to the 3D semigeostrophic system, and simple examples are provided by polynomials $S'(X,z)$.
The physical significance of this class of solutions is still unclear, but it appears to be related to the more general notion of ‘‘vertical slice models'' (see for example \cite{Cul2023} and references therein) and more investigation is needed to clarify this point.

Additional questions arise spontaneously from our investigations.
First, the characterization of \CS fronts as relatives of shocks in classical gas dynamics given here appears to be in apparent contrast to more classical constructions.
For example, Cullen \cite{Cul83} establishes that atmospheric fronts are advected by the fluid flow, and that a front continuously running from sea level to the tropopause cannot exist, but the solution to the Eady problem of Section \ref{subsec:x-waves} seems to violate both of these statements.
Therefore, further work is needed to fully understand the physical significance of a \CS front.


A further interesting question stemming from our studies on the Eady model concerns the physical meaning of curvature. In the context of 2D Euler flows, Napper \textit{et al.} \cite{NRR2023} pointed out that the regions where the solution is Riemannian and positively curved correspond to regions within the fluid flow dominated by vorticity and the presence of a vortex is expected. 
Further investigation is required to demonstrate a similar relationship in the context of buoyancy in the semigeostrophic model, although preliminary results presented here are encouraging.

\section{Acknowledgements}
The authors would like to thank M. Cullen and L. Napper for the useful discussions.
R.D. and G.O. were supported by the European Union’s Horizon 2020 research and innovation program under the Marie Skłodowska-Curie grant no 778010 IPaDEGAN. 
R.D. and G.O. thank the financial support of the project MMNLP (Mathematical Methods in Non Linear Physics) of the INFN.
R.D. and G.O. also gratefully acknowledge the auspices of the GNFM Section of INdAM under which part of this work was carried out. 
This work is part of R.D.'s dual PhD program Bicocca-Surrey.

\end{document}